\documentclass[prl, reprint, longbibliography, superscriptaddress, showpacs]{revtex4-1}
\usepackage{graphicx}
\usepackage{amssymb,amsmath}
\begin{document}
\widetext
\title{Nuclear resonant scattering from $^{193}$Ir as a probe of the electronic and magnetic properties of iridates} 
\author{P. Alexeev}
\affiliation{Deutsches Elektronen-Synchrotron DESY, Notkestra{\ss}e 85, 22607 Hamburg, Germany}
\affiliation{The Hamburg Centre for Ultrafast Imaging, Luruper Chaussee 149, 22761 Hamburg, Germany}
\author{O. Leupold}
\author{I. Sergueev}
\author{M. Herlitschke}
\affiliation{Deutsches Elektronen-Synchrotron DESY, Notkestra{\ss}e 85, 22607 Hamburg, Germany}
\author{D.F. McMorrow}
\author{R.S. Perry}
\author{E.C. Hunter}
\affiliation{London Centre for Nanotechnology and Department of Physics and Astronomy, University College London, Gower Street, London WC1E 6BT, United Kingdom}
\author{R. R\"{o}hlsberger}
\email{ralf.roehlsberger@desy.de}
\affiliation{Deutsches Elektronen-Synchrotron DESY, Notkestra{\ss}e 85, 22607 Hamburg, Germany}
\affiliation{The Hamburg Centre for Ultrafast Imaging, Luruper Chaussee 149, 22761 Hamburg, Germany}
\author{H.-C. Wille}
\email{hans.christian.wille@desy.de}
\affiliation{Deutsches Elektronen-Synchrotron DESY, Notkestra{\ss}e 85, 22607 Hamburg, Germany}
\date{\today}
\begin{abstract}
The high brilliance of the modern synchrotron radiation sources
facilitates experiments with high energy x-rays. In this Letter we
report on Nuclear Resonance Scattering at the 73 keV nuclear level in
$^{193}$Ir. The transitions between the hyperfine split levels show an
exceptionally large E2/M1 multi- polarity mixing ratio combined with
an increased sensitivity to certain changes in the hyperfine field
direction compared to non-mixing transitions. The method opens a new
way for probing local magnetic and electronic properties of correlated
materials containing iridium and provides novel insights into their
anisotropic magnetism. In particular, unexpected out-of-plane
components of magnetic hyperfine fields and non-zero electric field
gradients in Sr$_2$IrO$_4$ have been detected and attributed to the presence
of strong spin-orbit interaction. Due to the high, 62\,$\%$ natural
abundance of the $^{193}$Ir isotope, no isotopic enrichment of the samples
is required, qualifying the method for a broad range of applications.
\end{abstract}
%
%
\maketitle
%
%

There is burgeoning interest in understanding the physical properties
of systems which are simultaneously subject to 
strong spin-orbit coupling (SOC) and electron correlations, as
exemplified by recent studies which have revealed a range of novel
electronic and magnetic phases displayed by various 4d and 5d
transition metal oxides (TMOs) 
\cite{kimyk14, he15, CR94, kim09, bosegg13, kim08, moon08, don16, jack09, witcz14,
matth76, kimj12, kimyk16, dltor15}.

At one level, SOC introduces another competing energy scale, producing
unexpected electronic states.This is the case for the so-called
spin-orbit Mott insulator in iridate perovksites which would otherwise
be expected to be metallic in the absence of SOC. 
At another, more profound level, the SOC fully entangles spin and
orbital degrees of freedom such that the magnetic interactions acquire an anisotropic,
bond-directional nature -- the Kitaev interaction -- augmenting the
conventional isotropic Heisenberg term which dominates 3d systems
\cite{jack09}. The resulting Kitaev-Heisenberg model is proving to be
extremely rich displaying a plethora of topological quantum phases including
spin-liquids, superconductivity, etc., the exploration of which is in
its infancy \cite{witcz14, schaffer16}.
Further impetus for studying 4d and 5d TMOs stems from the rich
possibilities offered by nano-structuring  these materials, finding
potential applications as biosensors, spintronic devices, catalysts,
etc \cite{lin14, qiu15, hirsch99, fuj13}.

The iridate perovskites forming the Ruddlesden-Popper series of
compounds Sr$_{n+1}$Ir$_n$O$_{3n+1}$ play a central role in the 
evolution of the field of systems combining SOC and electron
correlations. Sr$_2$IrO$_4$ (n=1) was the first example of the new 
class of spin orbit Mott insulators which has attracted considerable
interest due to the similarities of its magnetism, and to a certain
extent its electronic structure,  to La$_2$CuO$_4$, the parent
compound of high-temperature superconductors. Indeed potassium doped
onto the surface of Sr$_2$IrO$_4$  has been shown to induce a d-wave
gap similar to that displayed by superconducting cuprates, although
definitive proof of superconductivity in the iridate perovskites has
not yet been produced. Sr$_2$I$_3$O$_7$ (n=2) is a marginal spin-orbit
Mott insulator, in the sense that it can be transformed to unusual
confined metallic phase (conducting in the ab plane only) for
pressures above ~55 GPa, although the details of key properties such
as the magnetism of the high-pressure phase are unknown 
\cite{CR94, kim09, bosegg13, moon08, yan15, kimyk16, wangsenth11}.

Indeed, revealing the nature of the electronic and magnetic
correlations in iridates presents certain challenges which need to be
overcome. These include the fact that the physics depends on a
hierarchy of  competing energy scales, requiring the characterisation
of  electronic and magnetic correlations over large ranges of energy
and length scales. Second, single crystals of novel materials are
often initially very small (in some cases no larger than 10 $\mu$m),
meaning that methods with high sensitivity have to be developed. X-ray 
resonant scattering, both elastic (REXS) and inelastic (RIXS), from
the Ir 4d electrons has proven to be especially useful, particularly
so as neutron techniques are less suitable due to the low sensitivity
of the technique and the high neutron absorption cross section of Ir. 

In this Letter we establish  synchrotron based nuclear resonance
scattering (NRS) on $^{193}$Ir at 73 keV as a  complementary probe to
REXS and RIXS for probing the electronic properties and magnetism of
iridates. The main advantages of NRS are its exquisite sensitivity to
the magnitude and direction of the electric and magnetic hyperfine
fields, rendering it uniquely capable of revealing subtle changes to
crystallographic and magnetic structures \cite{roe04, gerd99}. 
Moreover, the high photon energy of the $^{193}$Ir resonance
\cite{wagn70} opens the possibility of studying iridates
under extreme conditions of pressure, such as the insulator to metal
transition displayed by Sr$_3$Ir$_2$O$_7$ \cite{don16, ding16}.

Conventional M\"{o}ssbauer spectroscopy on Ir has
been performed several decades ago \cite{wagn70, perl69}, but did not
become widespread, because the preparation of radioactive sources was
notoriously difficult. NRS, on the other hand, does not require a
radioactive source. Moreover, the narrow collimation and small beam
size accessible at modern synchrotron radiation sources favor NRS
studies of nanostructures \cite{roeh01, stank07} and small samples at
extremely high pressures and temperatures \cite{roe04, gerd99, potapk13}.

Natural Ir occurs in two stable isotopes, $^{191}$Ir and $^{193}$Ir.
The 73\,keV transition in $^{193}$Ir with nuclear
spins of 3/2 and 1/2 of ground and excited state, respectively, is
most favorable for NRS studies due to the high, 62\% natural abundance of the
$^{193}$Ir isotope and a comparatively long natural lifetime of
8\,ns. The  experiments were conducted at the Dynamics Beamline P01 at PETRA III
(DESY, Hamburg) \cite{p01c}.
 
\begin{figure}
\includegraphics[scale=0.31]{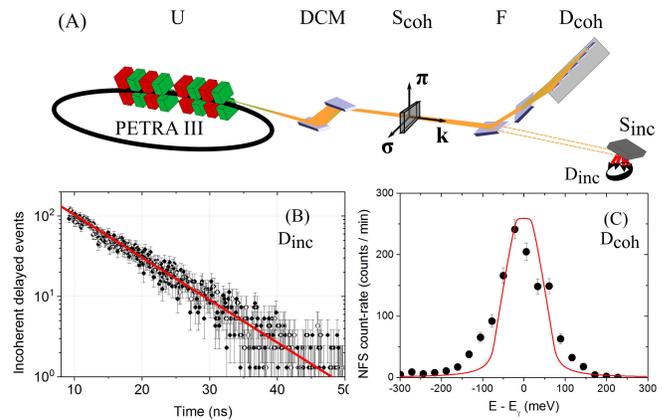}
\caption{\label{stp} (A): experimental setup with U - undulator
source, DCM - double crystal monochromator, F - filtering optics,
D$_{\rm{coh}}$ and D$_{\rm{inc}}$ - nuclear forward and nuclear
fluorescence APD detector, respectively, and S$_{\rm{coh}}$ and
S$_{\rm{inc}}$ - samples for forward and incoherent scattering
experiment, respectively. (B): time spectrum of delayed nuclear
fluorescence (black dots) and exponential decay with time constant
$\tau_0=8.4(2)$\,ns (red line). (C): spectrum of coherent delayed
events. The red line shows the instrumental function predicted by the dynamical theory of
x-ray diffraction.
}
\end{figure}

The storage ring was operated in 40-bunch top-up mode providing a
stable 100\,mA ring current. The experimental setup
(Fig. \ref{stp}\,(A)) included a double-crystal Si(3\,1\,1) monochromator
which reduced the energy bandwidth of the 73\,keV photons to about
8(1)\,eV. In order to reduce the energy bandwidth around the nuclear resonance
in $^{193}$Ir further, the energy bandwidth of photons transmitted
by the sample was filtered via Bragg reflections from two specially
designed Si crystals (F, Fig. \ref{stp}\,(A)). The first crystal
with an asymmetric (4\,4\,0) reflection collimates the beam for
matching the acceptance of the subsequent (6\,4\,2) reflection which
reduces the energy bandwidth to about 150\,meV \cite{SuppMat}. Tuning
of the photon energy to that of the nuclear resonance and
measurement of the lifetime of the excited state was performed by
monitoring delayed nuclear fluorescence from an Ir metal foil by a
large area avalanche photo diode (APD) detector (D$_{\rm{inc}}$,
Fig. \ref{stp}\,(A)). The nuclear forward scattering (NFS) was
detected by a fast detector array consisting of 16\,APDs (D$_{\rm{coh}}$,
Fig.\,\ref{stp}\,(A)) \cite{bar06, serg07, SuppMat}. The fast
detector array and very high bunch purity in the PETRA
storage ring enabled the counting of delayed photons as
early as 3\,ns after the excitation pulse with a time resolution of
about 0.6\,ns \cite{SuppMat}.\\

The nuclear resonance was found 3.211\,keV below the K-\,edge of Ir at 76.111 \,keV,
and its energy was determined to 72.90(8)\,keV. This value is in
good agreement with the frequently reported literature value of
73.0(5)\,keV \cite{nudat06}, though it is lower than the more precise value 73.045(5)\,keV
obtained in Ref. \cite{kishi05} from the measurement of internal
conversion. The reason for the latter is unclear as in both
measurements the maximum of the derivative of the edge absorption
curve was used as a reference. Furthermore, the reference value of the
edge used in \cite{kishi05} is 76.101 keV, which increases the
disagreement. Fitting the time spectrum of delayed nuclear
fluorescence with an exponential decay function (Fig. \ref{stp}\,(B)), 
we determined the natural lifetime to be 8.4(2)\,ns, in
accordance with the lifetime value of 8.8(2)\,ns reported
in Ref. \cite{nudat06} and slightly lower than the one reported in
\cite{kishi05} of 8.78 ns (no error given).

The corresponding resonance linewidth is 78(2)\,neV.
Using the NFS setup, we measured an instrumental function of the
filtering optics (Fig. \ref{stp}\,(C)). Its width of 158(8)\,meV
(FWHM) is close to that of 112\,meV (FWHM) predicted by dynamical
theory (Fig. \ref{stp}\,(C),\,red line). The broadening can be related
to the imperfections in the bulk silicon utilized for the crystals.

\begin{figure}
\includegraphics[scale=0.75]{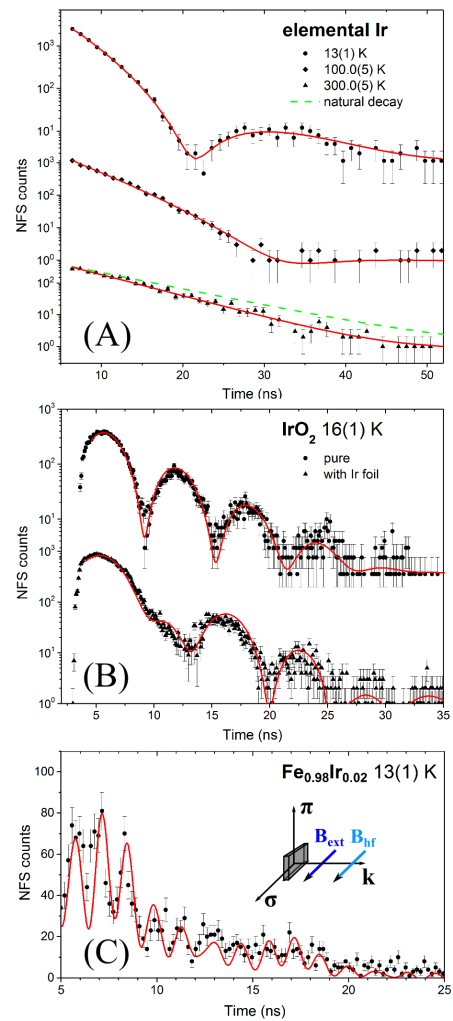}
\caption{\label{results} NFS time spectra of: (A) Ir foil, (B)
IrO$_2$ powder, (C) Fe$_{0.98}$Ir$_{0.02}$. Black markers show experimental data and the
red lines show fits by nuclear dynamical scattering theory. Green
dotted line in the lower graph of (A) shows the natural decay of the
73 keV state. For better visibility (C) is plotted in linear scale;
 the inset shows the scattering geometry, directions of external
 magnetic field $B_{ext}$ and hyperfine field $B_{hf}$.  }
\end{figure}

In order to demonstrate the feasibility of the technique we
performed NFS measurements on elemental Ir and on IrO$_2$. Both
materials have been studied earlier by conventional M\"{o}ssbauer
spectroscopy \cite{wagn83}; these are used here as references for
validation of data treatment routines in the time domain.
While elemental Ir shows a single resonance line, Ir in IrO$_2$
exhibits an Ir$^{4+}$ state with a pure electric hyperfine interaction \cite{wagn83}.\\
NRS time spectra of 100\,$\mu$m thick foil of elemental Ir have been
acquired in half an hour (Fig. \ref{results}\,(A)) at signal
countrates of about 7\,s$^{-1}$. We observed a shift of beating
minima to later times with increasing temperature due to the
decrease of Lamb-M\"{o}ssbauer factor (Fig. \ref{results}\,(A),
lower graphs). Since the temporal beating pattern can be fully
described as dynamical beats \cite{hann99}, hyperfine interactions
can be ruled out, in accordance with the cubic lattice and
paramagnetism of the elemental Ir \cite{kand13}. Fitting the
temperature dependence of the Lamb-M{\"o}ssbauer factor with the
Debye model \cite{Guetl}, we determined a Debye temperature of
309(30)K. This value is in good agreement with the literature
value of 335(13)\,K \cite{stein69}.\\
NFS time spectra of the IrO$_2$ powder sample are shown in Fig.
\ref{results}\,(B). Fitting of the experimental data (Fig. \ref{results}\,(B), upper
graph) was performed with the CONUSS software \cite{Sturh00,Sturh00}. 
To take the high mixing ratio of the E2/M1
multipole radiation into account \cite{Sturh94, wagn83} CONUSS had to
be extended. Special cases of the NRS theory for ferro-magnetic and
anti-ferromagnetic arrangements considering high mixing ratios are
rolled out in detail in the Supplemental Material \cite{SuppMat}. Where suitable, a
comparison to the simple M1 case in $^{57}$Fe is also given.

Fitting the data resulted in a quadrupole
splitting $\Delta E_Q = \frac{eQV_{zz}}{2}$ of 2.76(2)\,mm/s
(8.96(7)$\Gamma_0$) was obtained ($e$ is the elementary change, $Q$
is the quadrupole moment, $V_{zz}$ is the electric field gradient
(EFG) along the quantization axis). This value is in excellent agreement
with the value of 2.71(6)\,mm/s reported in Ref. \cite{atzm67}. We
obtained an axially symmetric EFG with
$V_{zz}$=1.71(1)$\cdot$10$^{18}$\,V/cm$^2$ which is two orders of
magnitude higher than in the isostructural 4d-RuO$_2$ reported in
Ref. \cite{bess14}. The EFG in IrO$_2$ is therefore mostly
determined by valence 5d-electrons because of: (i) three times lower
shielding of the Ir nucleus from the valence electrons than from the
surrounding ions \cite{ragh76, wagn83} and (ii) more elongated
5d-orbitals in IrO$_2$ providing a potentially higher EFG
\cite{Guetl}.
In order to measure the isomer shift of Ir$^{4+}$ in IrO$_2$, we
introduced an Ir metal foil as a single line reference absorber and
acquired a NFS time spectrum of the combined setup
(Fig. \ref{results}\,(B), lower graph). From the evaluation of this
dataset we obtained an isomer shift of -\,0.89(5)\,mm/s in IrO$_2$
relative to Ir metal, which is in good agreement with the value of
-\,0.93(1)\,mm/s reported in Ref. \cite{shen84}.

To develop the method for studies of magnetic materials, we measured
NFS from the ferromagnetic alloy Fe$_{0.98}$Ir$_{0.02}$ in an
external magnetic field of 0.53(5)\,T. Dilute alloys of
Fe$_{1-x}$Ir$_{x}$ ($x\leqslant 0.1$) show nearly pure magnetic
hyperfine interactions \cite{moessb71, salom74}, and the hyperfine
fields in these alloys are the highest for all known compounds with
d-elements \cite{rao75}. The large hyperfine fields lead to very
fast oscillations in the temporal beat patterns of NFS and therefore
provide the best benchmark of time resolution of the setup. The NFS
time spectrum of a 1.6\,mm thick sample of Fe$_{0.98}$Ir$_{0.02}$
exhibits extremely fast oscillations with a period of
$\approx1.5$\,ns (Fig. \ref{results}\,(C)).

Notably, despite the aforementioned high E2/M1 mixing ratio the
Fe$_{0.98}$Ir$_{0.02}$ NFS spectrum shows a very regular beating
pattern, significant for an (almost pure) two transition line
spectrum. At a first glance this is surprising as even the pure M1
case (e.g. for $^{57}$Fe) shows a more complicated spectrum. The
reason for this is that due to the E2/M1 mixing parameter, which value
is close the square root of 1/3, M1 and E2 transition amplitudes in
the mixed M1/E2 case can cancel each other for specific
transitions in $^{193}$Ir (for more details see Supplemental Material
\cite{SuppMat}, part C).
We refine the value of the hyperfine field to 133(1)\,T, which is in
good agreement with the value of 140(2)\,T reported for
Fe$_{0.973}$Ir$_{0.027}$ in Ref. \cite{wagn70}.

Having validated the NRS technique by studying relevant reference
samples, we applied it for exploring magnetism and electronic
properties of two iridates from the series of
{Sr$_{n+1}$Ir$_n$O$_{3n+1}$} Ruddlesden-Popper phases, i.e.
SrIrO$_3$ and Sr$_2$IrO$_4$.

\begin{figure}
\includegraphics[scale=0.75]{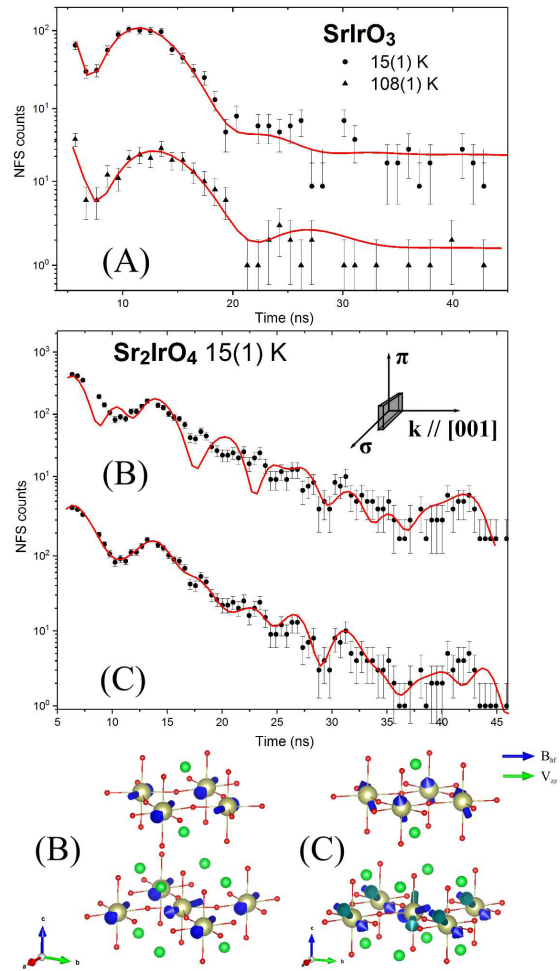}
\caption{\label{results2} (A): temperature dependent NFS time
spectra of SrIrO$_{3}$; (B): NFS time spectra of Sr$_2$IrO$_4$. The inset
shows the scattering geometry. The red lines are fits by nuclear
dynamical scattering theory: (B) assuming hyperfine fields in the
basal plane and (C) with hyperfine planes tilted from the basal
plane. The directions of the magnetic hyperfine fields $B_{hf}$ and
EFG quantization axes $V_{zz}$ for the corresponding model fit
in (B) and (C) are shown at the bottom .}
\end{figure}

SrIrO$_3$ is a paramagnetic metal with electronic structure
determined by both strong spin-orbit interaction and large
electronic correlation at temperatures higher than 50\,K
\cite{moon08, zhao08}. Due to its impact on the electronic
anisotropy and EFG resulting from it, the spin-orbit interaction in
SrIrO$_3$ can be studied by probing the quadrupole splitting of the
$^{193}$Ir nuclear levels \cite{inga64}. For a SrIrO$_3$ powder
sample we obtained a quadrupole splitting of 1.24(5)\,mm/s
(4.0(2)$\Gamma_0$) at 15\,K (Fig. \ref{results2}\,(A),\,upper graph),
in very good agreement with the value of 1.26\,mm/s measured at 4\,K
in Ref. \cite{shen84}. Magnetic hyperfine interactions can be ruled
out, in accordance with paramagnetism in this compound in the
temperature range under scope \cite{zhao08}. The quadrupole
splitting decreases with temperature and reaches a value of
1.08(5)\,mm/s (3.5(2)$\Gamma_0$) at 108\,K (Fig.
\ref{results2}\,(A),\,lower graph), which can be related to the
presence of a gap in the electronic ground state. To the best of our
knowledge, no change of Ir coordination symmetry is reported for the
temperature range investigated.
Therefore the temperature dependent change in quadrupole splitting can be
exclusively addressed to the thermal population of electronic levels,
supporting the evidence of semimetal-like electronic
band structure \cite{nie15} in SrIrO$_3$ and showing the decisive
impact of the large distortions in IrO$_6$ octahedra onto the
electronic structure in this compound.

It is the unique strength of the NRS technique to provide local
information on the magnitude and orientation of magnetic fields and
electric field gradients at the Ir sites \cite{SuppMat}. This allowed
us to gain new insights into the magnetic order of the Sr$_2$IrO$_4$
perovskite. Sr$_2$IrO$_4$ crystals have a form of platelets with
lateral size of 2x3 mm$^2$ and thickness of about 30-70\,$\mu$m; the
whole incident beam was accepted by the sample. The Sr$_2$IrO$_4$
sample was aligned with the ($0\,0\,1$) plane perpendicular to the
incident beam (inset Fig. \ref{results2}\,(B,C)) and five crystals have
been stacked in the same orientation along the beam in order to
increase the NFS signal \cite{SuppMat}.
EDX and magnetisation measurements suggest slight oxygen deficiency in
the Sr$_2$IrO$_4$ sample \cite{SuppMat}.

The temporal beat pattern in the time spectrum of Sr$_2$IrO$_4$ is
determined by both magnetic and electric hyperfine interactions
(Fig. \ref{results2}\,(B,C)). We obtained a hyperfine field of 24.2(2)\,T
which is in a very good agreement with the value of 24\,T reported
by M\"{o}ssbauer spectroscopy in Ref. \cite{wagn83}. Whilst NRS at
$^{193}$Ir shows high sensitivity to the orientation of hyperfine
fields \cite{SuppMat}, the model with in-plane hyperfine fields fails
to explain the measured time spectrum (Fig. \ref{results2}\,(B)).
Taking into account oxygen deficiency \cite{SuppMat} and associated
distortion of tetragonal symmetry \cite{yefe13}, the local symmetry
of Ir in Sr$_2$IrO$_4$ permits the existence of magnetic components
along the c-axis \cite{CR94,yefe13}. Introducing a 30$^{\rm{o}}$ tilting
angle of hyperfine fields to the a-b plane into the model fit
provides a very good statistical quality of the fit to the measured
time spectrum (Fig. \ref{results2}\,(C)), supporting the existence
of out-of-plane components of the magnetic field at the Ir sites
which was not observed before. One has to note that the direction of
hyperfine field and magnetic moment do not need to coincide
\cite{schuenem00}. Especially, the effect is expected in the
presence of significant orbital field contribution and interaction
with the lattice as reported in Ref. \cite{gret16}. An enhanced
electronegativity of Ir also favors high covalency of Ir-O bonds,
reducing the Fermi contact field and increasing the orbital field
contribution to the hyperfine field \cite{oost69, henn67, herlit14}.
In accordance with this statement, a strong anisotropy in electronic
$g$-factors in Sr$_2$IrO$_4$  was observed by ESR in Ref. \cite{bogd15}.\\
Considering the electric field at the Ir nuclei in Sr$_2$IrO$_4$, we
observe an axially symmetric EFG with a magnitude of
$1.1(1)\cdot10^{18}$\,V/cm$^2$ in the [$0\,0\,1$] direction. The
presence of an EFG is reasonable in view of distortion of the
IrO$_6$ octahedra \cite{CR94}, oxygen deficiency, and the evidence
of a non-zero EFG in the isostructural Sr$_2$RuO$_{4}$
\cite{ishid97}. The non-zero EFG in Sr$_2$IrO$_4$ and the
out-of-plane components of the magnetic hyperfine field found here
might be addressed to the non-zero angular momentum
of the outer electrons, arising from the reduced symmetry of the IrO$_6$ octahedra;
temperature dependent measurements can provide further information on
the origin of this phenomenon.

In conclusion, we have established Nuclear Resonance Scattering at
the 72.90(8)\,keV level in $^{193}$Ir as a new synchrotron-based
technique for the studies of magnetism and electronic properties of
iridates. A huge 133(1)\,T hyperfine field in dilute Fe$_{0.98}$Ir$_{0.02}$
alloy has been detected via NRS. Moreover, we found a thermally
induced decrease of the electric field gradient across the Ir nuclei
in SrIrO$_{3}$ and observed a non-zero EFG and tilting of hyperfine
fields from the basal plane in Sr$_2$IrO$_4$ that should stimulate further investigations
to relate structural and electronic properties in the iridates. All samples contained
$^{193}$Ir in its natural abundance; no preparation of radioactive
sources is required and no line broadening due to the source is
present. NRS at $^{193}$Ir is sensitive to dilute systems and spin
structures, providing a valuable input for studies to relate magnetism and
spin-orbit interactions in iridates, e.g. in strong magnetic fields
\cite{kim09, perl69}, or under confinement in nanomaterials
\cite{hirsch99, lin14, fuj13} and heterostructures \cite{nich16}.
The oxidation state of iridium and crystal fields at Ir ions can be
tracked via measurements of isomer shift and quadrupole interactions
at the Ir nucleus, respectively.
\begin{acknowledgments}
We acknowledge support of the Helmholtz association via project
oriented funds. The PETRA machine operation group is gratefully
acknowledged for establishing a beam cleaning procedure and
maintaining high bunch purity. Wolfgang Sturhahn is greatly
acknowledged for extending the CONUSS software for calculations of
mixed multipole radiation. The authors are thankful to Hlynur
Gretarsson, Christian Donnerer, and Rapha\"{e}l P. Hermann for
fruitful discussions on the physics of iridates. We thank
Manfred Spiwek and Frank-Uwe Dill for the preparation of the silicon
crystals and setup at the beamline. Thomas F. Keller (DESY NanoLab)
is acknowledged for EDX measurements on Sr$_2$IrO$_4$. Work at UCL
was supported by the Engineering and Physical Sciences Research Council
(grants EP/N027671/1 and EP/N034694/1).
\end{acknowledgments}
%
%
%
%
%

\begin{thebibliography}{65}%
\makeatletter
\providecommand \@ifxundefined [1]{%
 \@ifx{#1\undefined}
}%
\providecommand \@ifnum [1]{%
 \ifnum #1\expandafter \@firstoftwo
 \else \expandafter \@secondoftwo
 \fi
}%
\providecommand \@ifx [1]{%
 \ifx #1\expandafter \@firstoftwo
 \else \expandafter \@secondoftwo
 \fi
}%
\providecommand \natexlab [1]{#1}%
\providecommand \enquote  [1]{``#1''}%
\providecommand \bibnamefont  [1]{#1}%
\providecommand \bibfnamefont [1]{#1}%
\providecommand \citenamefont [1]{#1}%
\providecommand \href@noop [0]{\@secondoftwo}%
\providecommand \href [0]{\begingroup \@sanitize@url \@href}%
\providecommand \@href[1]{\@@startlink{#1}\@@href}%
\providecommand \@@href[1]{\endgroup#1\@@endlink}%
\providecommand \@sanitize@url [0]{\catcode `\\12\catcode `\$12\catcode
  `\&12\catcode `\#12\catcode `\^12\catcode `\_12\catcode `\%12\relax}%
\providecommand \@@startlink[1]{}%
\providecommand \@@endlink[0]{}%
\providecommand \url  [0]{\begingroup\@sanitize@url \@url }%
\providecommand \@url [1]{\endgroup\@href {#1}{\urlprefix }}%
\providecommand \urlprefix  [0]{URL }%
\providecommand \Eprint [0]{\href }%
\providecommand \doibase [0]{http://dx.doi.org/}%
\providecommand \selectlanguage [0]{\@gobble}%
\providecommand \bibinfo  [0]{\@secondoftwo}%
\providecommand \bibfield  [0]{\@secondoftwo}%
\providecommand \translation [1]{[#1]}%
\providecommand \BibitemOpen [0]{}%
\providecommand \bibitemStop [0]{}%
\providecommand \bibitemNoStop [0]{.\EOS\space}%
\providecommand \EOS [0]{\spacefactor3000\relax}%
\providecommand \BibitemShut  [1]{\csname bibitem#1\endcsname}%
\let\auto@bib@innerbib\@empty
\bibitem [{\citenamefont {Kim}\ \emph {et~al.}(2014)\citenamefont {Kim},
  \citenamefont {Krupin}, \citenamefont {Denlinger}, \citenamefont {Bostwick},
  \citenamefont {Rotenberg}, \citenamefont {Zhao}, \citenamefont {Mitchell},
  \citenamefont {Allen},\ and\ \citenamefont {Kim}}]{kimyk14}%
  \BibitemOpen
  \bibfield  {author} {\bibinfo {author} {\bibfnamefont {Y.~K.}\ \bibnamefont
  {Kim}}, \bibinfo {author} {\bibfnamefont {O.}~\bibnamefont {Krupin}},
  \bibinfo {author} {\bibfnamefont {J.~D.}\ \bibnamefont {Denlinger}}, \bibinfo
  {author} {\bibfnamefont {A.}~\bibnamefont {Bostwick}}, \bibinfo {author}
  {\bibfnamefont {E.}~\bibnamefont {Rotenberg}}, \bibinfo {author}
  {\bibfnamefont {Q.}~\bibnamefont {Zhao}}, \bibinfo {author} {\bibfnamefont
  {J.~F.}\ \bibnamefont {Mitchell}}, \bibinfo {author} {\bibfnamefont {J.~W.}\
  \bibnamefont {Allen}}, \ and\ \bibinfo {author} {\bibfnamefont {B.~J.}\
  \bibnamefont {Kim}},\ }\bibfield  {title} {\enquote {\bibinfo {title} {{Fermi
  arcs in a doped pseudospin-1/2 Heisenberg antiferromagnet}},}\ }\href
  {\doibase 10.1126/science.1251151} {\bibfield  {journal} {\bibinfo  {journal}
  {Science}\ }\textbf {\bibinfo {volume} {345}},\ \bibinfo {pages} {187--190}
  (\bibinfo {year} {2014})}\BibitemShut {NoStop}%
\bibitem [{\citenamefont {He}\ \emph {et~al.}(2015)\citenamefont {He},
  \citenamefont {Hafiz}, \citenamefont {Mion}, \citenamefont {Hogan},
  \citenamefont {Dhital}, \citenamefont {Chen}, \citenamefont {Lin},
  \citenamefont {Hashimoto}, \citenamefont {Lu}, \citenamefont {Zhang},
  \citenamefont {Markiewicz}, \citenamefont {Bansil}, \citenamefont {Wilson},\
  and\ \citenamefont {He}}]{he15}%
  \BibitemOpen
  \bibfield  {author} {\bibinfo {author} {\bibfnamefont {J.}~\bibnamefont
  {He}}, \bibinfo {author} {\bibfnamefont {H.}~\bibnamefont {Hafiz}}, \bibinfo
  {author} {\bibfnamefont {Th.~R.}\ \bibnamefont {Mion}}, \bibinfo {author}
  {\bibfnamefont {T.}~\bibnamefont {Hogan}}, \bibinfo {author} {\bibfnamefont
  {C.}~\bibnamefont {Dhital}}, \bibinfo {author} {\bibfnamefont
  {X.}~\bibnamefont {Chen}}, \bibinfo {author} {\bibfnamefont {Q.}~\bibnamefont
  {Lin}}, \bibinfo {author} {\bibfnamefont {M.}~\bibnamefont {Hashimoto}},
  \bibinfo {author} {\bibfnamefont {D.~H.}\ \bibnamefont {Lu}}, \bibinfo
  {author} {\bibfnamefont {Y.}~\bibnamefont {Zhang}}, \bibinfo {author}
  {\bibfnamefont {R.~S.}\ \bibnamefont {Markiewicz}}, \bibinfo {author}
  {\bibfnamefont {A.}~\bibnamefont {Bansil}}, \bibinfo {author} {\bibfnamefont
  {S.~D.}\ \bibnamefont {Wilson}}, \ and\ \bibinfo {author} {\bibfnamefont
  {Rui-Hua}\ \bibnamefont {He}},\ }\bibfield  {title} {\enquote {\bibinfo
  {title} {{F}ermi {Arcs} vs. {F}ermi {P}ockets in {E}lectron-doped
  {P}erovskite {I}ridates},}\ }\href {\doibase 10.1038/srep08533} {\bibfield
  {journal} {\bibinfo  {journal} {Sci. Rep.}\ }\textbf {\bibinfo {volume}
  {5}},\ \bibinfo {pages} {8533} (\bibinfo {year} {2015})}\BibitemShut
  {NoStop}%
\bibitem [{\citenamefont {Crawford}\ \emph {et~al.}(1994)\citenamefont
  {Crawford}, \citenamefont {Subramanian}, \citenamefont {Harlow},
  \citenamefont {Fernandez-Baca}, \citenamefont {Wang},\ and\ \citenamefont
  {Johnston}}]{CR94}%
  \BibitemOpen
  \bibfield  {author} {\bibinfo {author} {\bibfnamefont {M.~K.}\ \bibnamefont
  {Crawford}}, \bibinfo {author} {\bibfnamefont {M.~A.}\ \bibnamefont
  {Subramanian}}, \bibinfo {author} {\bibfnamefont {R.~L.}\ \bibnamefont
  {Harlow}}, \bibinfo {author} {\bibfnamefont {J.~A.}\ \bibnamefont
  {Fernandez-Baca}}, \bibinfo {author} {\bibfnamefont {Z.~R.}\ \bibnamefont
  {Wang}}, \ and\ \bibinfo {author} {\bibfnamefont {D.~C.}\ \bibnamefont
  {Johnston}},\ }\bibfield  {title} {\enquote {\bibinfo {title} {{Structural
  and magnetic studies of {Sr$_{2}$IrO$_{4}$}}},}\ }\href {\doibase
  10.1103/PhysRevB.49.9198} {\bibfield  {journal} {\bibinfo  {journal} {Phys.
  Rev. B}\ }\textbf {\bibinfo {volume} {49}},\ \bibinfo {pages} {9198--9201}
  (\bibinfo {year} {1994})}\BibitemShut {NoStop}%
\bibitem [{\citenamefont {Kim}\ \emph {et~al.}(2009)\citenamefont {Kim},
  \citenamefont {Ohsumi}, \citenamefont {Komesu}, \citenamefont {Sakai},
  \citenamefont {Morita}, \citenamefont {Takagi},\ and\ \citenamefont
  {Arima}}]{kim09}%
  \BibitemOpen
  \bibfield  {author} {\bibinfo {author} {\bibfnamefont {B.~J.}\ \bibnamefont
  {Kim}}, \bibinfo {author} {\bibfnamefont {H.}~\bibnamefont {Ohsumi}},
  \bibinfo {author} {\bibfnamefont {T.}~\bibnamefont {Komesu}}, \bibinfo
  {author} {\bibfnamefont {S.}~\bibnamefont {Sakai}}, \bibinfo {author}
  {\bibfnamefont {T.}~\bibnamefont {Morita}}, \bibinfo {author} {\bibfnamefont
  {H.}~\bibnamefont {Takagi}}, \ and\ \bibinfo {author} {\bibfnamefont
  {T.}~\bibnamefont {Arima}},\ }\bibfield  {title} {\enquote {\bibinfo {title}
  {Phase-sensitive observation of a spin-orbital mott state in
  {Sr$_{2}$IrO$_{4}$}},}\ }\href {\doibase 10.1126/science.1167106} {\bibfield
  {journal} {\bibinfo  {journal} {Science}\ }\textbf {\bibinfo {volume}
  {323}},\ \bibinfo {pages} {1329--1332} (\bibinfo {year} {2009})}\BibitemShut
  {NoStop}%
\bibitem [{\citenamefont {Boseggia}\ \emph {et~al.}(2013)\citenamefont
  {Boseggia}, \citenamefont {Walker}, \citenamefont {Vale}, \citenamefont
  {Springell}, \citenamefont {Feng}, \citenamefont {Perry}, \citenamefont
  {Sala}, \citenamefont {{R{\o}nnow}}, \citenamefont {Collins},\ and\
  \citenamefont {McMorrow}}]{bosegg13}%
  \BibitemOpen
  \bibfield  {author} {\bibinfo {author} {\bibfnamefont {S.}~\bibnamefont
  {Boseggia}}, \bibinfo {author} {\bibfnamefont {H.C.}\ \bibnamefont {Walker}},
  \bibinfo {author} {\bibfnamefont {J.}~\bibnamefont {Vale}}, \bibinfo {author}
  {\bibfnamefont {R.}~\bibnamefont {Springell}}, \bibinfo {author}
  {\bibfnamefont {Z.}~\bibnamefont {Feng}}, \bibinfo {author} {\bibfnamefont
  {R.S.}\ \bibnamefont {Perry}}, \bibinfo {author} {\bibfnamefont {M.M.}\
  \bibnamefont {Sala}}, \bibinfo {author} {\bibfnamefont {H.M.}\ \bibnamefont
  {{R{\o}nnow}}}, \bibinfo {author} {\bibfnamefont {S.P.}\ \bibnamefont
  {Collins}}, \ and\ \bibinfo {author} {\bibfnamefont {D.F.}\ \bibnamefont
  {McMorrow}},\ }\bibfield  {title} {\enquote {\bibinfo {title} {Locking of
  iridium magnetic moments to the correlated rotation of oxygen octahedra in
  {Sr$_{2}$IrO$_{4}$} revealed by x-ray resonant scattering},}\ }\href
  {\doibase 10.1088/0953-8984/25/42/422202} {\bibfield  {journal} {\bibinfo
  {journal} {J. Phys. Condens. Matter}\ }\textbf {\bibinfo {volume} {25}},\
  \bibinfo {pages} {422202} (\bibinfo {year} {2013})}\BibitemShut {NoStop}%
\bibitem [{\citenamefont {Kim}\ \emph {et~al.}(2008)\citenamefont {Kim},
  \citenamefont {Jin}, \citenamefont {Moon}, \citenamefont {Kim}, \citenamefont
  {Park}, \citenamefont {Leem}, \citenamefont {Yu}, \citenamefont {Noh},
  \citenamefont {Kim}, \citenamefont {Oh}, \citenamefont {Park}, \citenamefont
  {Durairaj}, \citenamefont {Cao},\ and\ \citenamefont {Rotenberg}}]{kim08}%
  \BibitemOpen
  \bibfield  {author} {\bibinfo {author} {\bibfnamefont {B.~J.}\ \bibnamefont
  {Kim}}, \bibinfo {author} {\bibfnamefont {Hosub}\ \bibnamefont {Jin}},
  \bibinfo {author} {\bibfnamefont {S.~J.}\ \bibnamefont {Moon}}, \bibinfo
  {author} {\bibfnamefont {J.-Y.}\ \bibnamefont {Kim}}, \bibinfo {author}
  {\bibfnamefont {B.-G.}\ \bibnamefont {Park}}, \bibinfo {author}
  {\bibfnamefont {C.~S.}\ \bibnamefont {Leem}}, \bibinfo {author}
  {\bibfnamefont {Jaejun}\ \bibnamefont {Yu}}, \bibinfo {author} {\bibfnamefont
  {T.~W.}\ \bibnamefont {Noh}}, \bibinfo {author} {\bibfnamefont
  {C.}~\bibnamefont {Kim}}, \bibinfo {author} {\bibfnamefont {S.-J.}\
  \bibnamefont {Oh}}, \bibinfo {author} {\bibfnamefont {J.-H.}\ \bibnamefont
  {Park}}, \bibinfo {author} {\bibfnamefont {V.}~\bibnamefont {Durairaj}},
  \bibinfo {author} {\bibfnamefont {G.}~\bibnamefont {Cao}}, \ and\ \bibinfo
  {author} {\bibfnamefont {E.}~\bibnamefont {Rotenberg}},\ }\bibfield  {title}
  {\enquote {\bibinfo {title} {{Novel ${J}_{eff}=1/2$ Mott State Induced by
  Relativistic Spin-Orbit Coupling in {Sr$_{2}$IrO$_{4}$}}},}\ }\href {\doibase
  10.1103/PhysRevLett.101.076402} {\bibfield  {journal} {\bibinfo  {journal}
  {Phys. Rev. Lett.}\ }\textbf {\bibinfo {volume} {101}},\ \bibinfo {pages}
  {076402} (\bibinfo {year} {2008})}\BibitemShut {NoStop}%
\bibitem [{\citenamefont {Moon}\ \emph {et~al.}(2008)\citenamefont {Moon},
  \citenamefont {Jin}, \citenamefont {Kim}, \citenamefont {Choi}, \citenamefont
  {Lee}, \citenamefont {Yu}, \citenamefont {Cao}, \citenamefont {Sumi},
  \citenamefont {Funakubo}, \citenamefont {Bernhard},\ and\ \citenamefont
  {Noh}}]{moon08}%
  \BibitemOpen
  \bibfield  {author} {\bibinfo {author} {\bibfnamefont {S.~J.}\ \bibnamefont
  {Moon}}, \bibinfo {author} {\bibfnamefont {H.}~\bibnamefont {Jin}}, \bibinfo
  {author} {\bibfnamefont {K.~W.}\ \bibnamefont {Kim}}, \bibinfo {author}
  {\bibfnamefont {W.~S.}\ \bibnamefont {Choi}}, \bibinfo {author}
  {\bibfnamefont {Y.~S.}\ \bibnamefont {Lee}}, \bibinfo {author} {\bibfnamefont
  {J.}~\bibnamefont {Yu}}, \bibinfo {author} {\bibfnamefont {G.}~\bibnamefont
  {Cao}}, \bibinfo {author} {\bibfnamefont {A.}~\bibnamefont {Sumi}}, \bibinfo
  {author} {\bibfnamefont {H.}~\bibnamefont {Funakubo}}, \bibinfo {author}
  {\bibfnamefont {C.}~\bibnamefont {Bernhard}}, \ and\ \bibinfo {author}
  {\bibfnamefont {T.~W.}\ \bibnamefont {Noh}},\ }\bibfield  {title} {\enquote
  {\bibinfo {title} {{Dimensionality-Controlled Insulator-Metal Transition and
  Correlated Metallic State in $5d$ Transition Metal Oxides
  {Sr$_{n+1}$Ir$_n$O$_{3n+1}$} ($n=1$, 2, and $\infty$)}},}\ }\href {\doibase
  10.1103/PhysRevLett.101.226402} {\bibfield  {journal} {\bibinfo  {journal}
  {Phys. Rev. Lett.}\ }\textbf {\bibinfo {volume} {101}},\ \bibinfo {pages}
  {226402} (\bibinfo {year} {2008})}\BibitemShut {NoStop}%
\bibitem [{\citenamefont {Donnerer}\ \emph {et~al.}(2016)\citenamefont
  {Donnerer}, \citenamefont {Feng}, \citenamefont {Vale}, \citenamefont
  {Andreev}, \citenamefont {Solovyev}, \citenamefont {Hunter}, \citenamefont
  {Hanfland}, \citenamefont {Perry}, \citenamefont {{R{\o}nnow}}, \citenamefont
  {McMahon}, \citenamefont {Mazurenko},\ and\ \citenamefont
  {McMorrow}}]{don16}%
  \BibitemOpen
  \bibfield  {author} {\bibinfo {author} {\bibfnamefont {C.}~\bibnamefont
  {Donnerer}}, \bibinfo {author} {\bibfnamefont {Z.}~\bibnamefont {Feng}},
  \bibinfo {author} {\bibfnamefont {J.~G.}\ \bibnamefont {Vale}}, \bibinfo
  {author} {\bibfnamefont {S.~N.}\ \bibnamefont {Andreev}}, \bibinfo {author}
  {\bibfnamefont {I.~V.}\ \bibnamefont {Solovyev}}, \bibinfo {author}
  {\bibfnamefont {E.~C.}\ \bibnamefont {Hunter}}, \bibinfo {author}
  {\bibfnamefont {M.}~\bibnamefont {Hanfland}}, \bibinfo {author}
  {\bibfnamefont {R.~S.}\ \bibnamefont {Perry}}, \bibinfo {author}
  {\bibfnamefont {H.~M.}\ \bibnamefont {{R{\o}nnow}}}, \bibinfo {author}
  {\bibfnamefont {M.~I.}\ \bibnamefont {McMahon}}, \bibinfo {author}
  {\bibfnamefont {V.~V.}\ \bibnamefont {Mazurenko}}, \ and\ \bibinfo {author}
  {\bibfnamefont {D.~F.}\ \bibnamefont {McMorrow}},\ }\bibfield  {title}
  {\enquote {\bibinfo {title} {Pressure dependence of the structure and
  electronic properties of {Sr$_{3}$Ir$_{2}$O$_{7}$}},}\ }\href {\doibase
  10.1103/PhysRevB.93.174118} {\bibfield  {journal} {\bibinfo  {journal} {Phys.
  Rev. B}\ }\textbf {\bibinfo {volume} {93}},\ \bibinfo {pages} {174118}
  (\bibinfo {year} {2016})}\BibitemShut {NoStop}%
\bibitem [{\citenamefont {Jackeli}\ and\ \citenamefont
  {Khaliullin}(2009)}]{jack09}%
  \BibitemOpen
  \bibfield  {author} {\bibinfo {author} {\bibfnamefont {G.}~\bibnamefont
  {Jackeli}}\ and\ \bibinfo {author} {\bibfnamefont {G.}~\bibnamefont
  {Khaliullin}},\ }\bibfield  {title} {\enquote {\bibinfo {title} {Mott
  insulators in the strong spin-orbit coupling limit: From {Heisenberg} to a
  quantum compass and {Kitaev} models},}\ }\href {\doibase
  10.1103/PhysRevLett.102.017205} {\bibfield  {journal} {\bibinfo  {journal}
  {Phys. Rev. Lett.}\ }\textbf {\bibinfo {volume} {102}},\ \bibinfo {pages}
  {017205} (\bibinfo {year} {2009})}\BibitemShut {NoStop}%
\bibitem [{\citenamefont {Witczak-Krempa}\ \emph {et~al.}(2014)\citenamefont
  {Witczak-Krempa}, \citenamefont {Chen}, \citenamefont {Kim},\ and\
  \citenamefont {Balents}}]{witcz14}%
  \BibitemOpen
  \bibfield  {author} {\bibinfo {author} {\bibfnamefont {W.}~\bibnamefont
  {Witczak-Krempa}}, \bibinfo {author} {\bibfnamefont {G.}~\bibnamefont
  {Chen}}, \bibinfo {author} {\bibfnamefont {Y.B.}\ \bibnamefont {Kim}}, \ and\
  \bibinfo {author} {\bibfnamefont {L.}~\bibnamefont {Balents}},\ }\bibfield
  {title} {\enquote {\bibinfo {title} {{Correlated Quantum Phenomena in the
  Strong Spin-Orbit Regime}},}\ }\href {\doibase
  10.1146/annurev-conmatphys-020911-125138} {\bibfield  {journal} {\bibinfo
  {journal} {Annu. Rev. Condens. Matter Phys.}\ }\textbf {\bibinfo {volume}
  {5}},\ \bibinfo {pages} {57--82} (\bibinfo {year} {2014})}\BibitemShut
  {NoStop}%
\bibitem [{\citenamefont {Mattheiss}(1976)}]{matth76}%
  \BibitemOpen
  \bibfield  {author} {\bibinfo {author} {\bibfnamefont {L.~F.}\ \bibnamefont
  {Mattheiss}},\ }\bibfield  {title} {\enquote {\bibinfo {title} {{Electronic
  structure of RuO$_{2}$, OsO$_{2}$, and IrO$_{2}$}},}\ }\href {\doibase
  10.1103/PhysRevB.13.2433} {\bibfield  {journal} {\bibinfo  {journal} {Phys.
  Rev. B}\ }\textbf {\bibinfo {volume} {13}},\ \bibinfo {pages} {2433--2450}
  (\bibinfo {year} {1976})}\BibitemShut {NoStop}%
\bibitem [{\citenamefont {Kim}\ \emph {et~al.}(2012)\citenamefont {Kim},
  \citenamefont {Casa}, \citenamefont {Upton}, \citenamefont {Gog},
  \citenamefont {Kim}, \citenamefont {Mitchell}, \citenamefont {van
  Veenendaal}, \citenamefont {Daghofer}, \citenamefont {van~den Brink},
  \citenamefont {Khaliullin},\ and\ \citenamefont {Kim}}]{kimj12}%
  \BibitemOpen
  \bibfield  {author} {\bibinfo {author} {\bibfnamefont {J.}~\bibnamefont
  {Kim}}, \bibinfo {author} {\bibfnamefont {D.}~\bibnamefont {Casa}}, \bibinfo
  {author} {\bibfnamefont {M.~H.}\ \bibnamefont {Upton}}, \bibinfo {author}
  {\bibfnamefont {T.}~\bibnamefont {Gog}}, \bibinfo {author} {\bibfnamefont
  {Y.-J.}\ \bibnamefont {Kim}}, \bibinfo {author} {\bibfnamefont {J.~F.}\
  \bibnamefont {Mitchell}}, \bibinfo {author} {\bibfnamefont {M.}~\bibnamefont
  {van Veenendaal}}, \bibinfo {author} {\bibfnamefont {M.}~\bibnamefont
  {Daghofer}}, \bibinfo {author} {\bibfnamefont {J.}~\bibnamefont {van~den
  Brink}}, \bibinfo {author} {\bibfnamefont {G.}~\bibnamefont {Khaliullin}}, \
  and\ \bibinfo {author} {\bibfnamefont {B.~J.}\ \bibnamefont {Kim}},\
  }\bibfield  {title} {\enquote {\bibinfo {title} {{Magnetic Excitation Spectra
  of {Sr$_{2}$IrO$_{4}$} Probed by Resonant Inelastic X-Ray Scattering:
  Establishing Links to Cuprate Superconductors}},}\ }\href {\doibase
  10.1103/PhysRevLett.108.177003} {\bibfield  {journal} {\bibinfo  {journal}
  {Phys. Rev. Lett.}\ }\textbf {\bibinfo {volume} {108}},\ \bibinfo {pages}
  {177003} (\bibinfo {year} {2012})}\BibitemShut {NoStop}%
\bibitem [{\citenamefont {Kim}\ \emph {et~al.}(2016)\citenamefont {Kim},
  \citenamefont {Sung}, \citenamefont {Denlinger},\ and\ \citenamefont
  {Kim}}]{kimyk16}%
  \BibitemOpen
  \bibfield  {author} {\bibinfo {author} {\bibfnamefont {Y.~K.}\ \bibnamefont
  {Kim}}, \bibinfo {author} {\bibfnamefont {N.~H.}\ \bibnamefont {Sung}},
  \bibinfo {author} {\bibfnamefont {J.~D.}\ \bibnamefont {Denlinger}}, \ and\
  \bibinfo {author} {\bibfnamefont {B.~J.}\ \bibnamefont {Kim}},\ }\bibfield
  {title} {\enquote {\bibinfo {title} {Observation of a d-wave gap in
  electron-doped {Sr$_{2}$IrO$_{4}$}},}\ }\href {\doibase 10.1038/nphys3503}
  {\bibfield  {journal} {\bibinfo  {journal} {Nature Phys.}\ }\textbf {\bibinfo
  {volume} {12}},\ \bibinfo {pages} {37--41} (\bibinfo {year}
  {2016})}\BibitemShut {NoStop}%
\bibitem [{\citenamefont {de~la Torre}\ \emph {et~al.}(2015)\citenamefont
  {de~la Torre}, \citenamefont {McKeown~Walker}, \citenamefont {Bruno},
  \citenamefont {Ricc\'o}, \citenamefont {Wang}, \citenamefont
  {Gutierrez~Lezama}, \citenamefont {Scheerer}, \citenamefont {Giriat},
  \citenamefont {Jaccard}, \citenamefont {Berthod}, \citenamefont {Kim},
  \citenamefont {Hoesch}, \citenamefont {Hunter}, \citenamefont {Perry},
  \citenamefont {Tamai},\ and\ \citenamefont {Baumberger}}]{dltor15}%
  \BibitemOpen
  \bibfield  {author} {\bibinfo {author} {\bibfnamefont {A.}~\bibnamefont
  {de~la Torre}}, \bibinfo {author} {\bibfnamefont {S.}~\bibnamefont
  {McKeown~Walker}}, \bibinfo {author} {\bibfnamefont {F.~Y.}\ \bibnamefont
  {Bruno}}, \bibinfo {author} {\bibfnamefont {S.}~\bibnamefont {Ricc\'o}},
  \bibinfo {author} {\bibfnamefont {Z.}~\bibnamefont {Wang}}, \bibinfo {author}
  {\bibfnamefont {I.}~\bibnamefont {Gutierrez~Lezama}}, \bibinfo {author}
  {\bibfnamefont {G.}~\bibnamefont {Scheerer}}, \bibinfo {author}
  {\bibfnamefont {G.}~\bibnamefont {Giriat}}, \bibinfo {author} {\bibfnamefont
  {D.}~\bibnamefont {Jaccard}}, \bibinfo {author} {\bibfnamefont
  {C.}~\bibnamefont {Berthod}}, \bibinfo {author} {\bibfnamefont {T.~K.}\
  \bibnamefont {Kim}}, \bibinfo {author} {\bibfnamefont {M.}~\bibnamefont
  {Hoesch}}, \bibinfo {author} {\bibfnamefont {E.~C.}\ \bibnamefont {Hunter}},
  \bibinfo {author} {\bibfnamefont {R.~S.}\ \bibnamefont {Perry}}, \bibinfo
  {author} {\bibfnamefont {A.}~\bibnamefont {Tamai}}, \ and\ \bibinfo {author}
  {\bibfnamefont {F.}~\bibnamefont {Baumberger}},\ }\bibfield  {title}
  {\enquote {\bibinfo {title} {Collapse of the mott gap and emergence of a
  nodal liquid in lightly doped {Sr$_{2}$IrO$_{4}$}},}\ }\href {\doibase
  10.1103/PhysRevLett.115.176402} {\bibfield  {journal} {\bibinfo  {journal}
  {Phys. Rev. Lett.}\ }\textbf {\bibinfo {volume} {115}},\ \bibinfo {pages}
  {176402} (\bibinfo {year} {2015})}\BibitemShut {NoStop}%
\bibitem [{\citenamefont {Schaffer}\ \emph {et~al.}(2016)\citenamefont
  {Schaffer}, \citenamefont {Kin-Ho~Lee}, \citenamefont {Yang},\ and\
  \citenamefont {Kim}}]{schaffer16}%
  \BibitemOpen
  \bibfield  {author} {\bibinfo {author} {\bibfnamefont {R.}~\bibnamefont
  {Schaffer}}, \bibinfo {author} {\bibfnamefont {E.}~\bibnamefont
  {Kin-Ho~Lee}}, \bibinfo {author} {\bibfnamefont {B.-J.}\ \bibnamefont
  {Yang}}, \ and\ \bibinfo {author} {\bibfnamefont {Y.~B.}\ \bibnamefont
  {Kim}},\ }\bibfield  {title} {\enquote {\bibinfo {title} {Recent progress on
  correlated electron systems with strong spin–orbit coupling},}\ }\href
  {http://stacks.iop.org/0034-4885/79/i=9/a=094504} {\bibfield  {journal}
  {\bibinfo  {journal} {Reports on Progress in Physics}\ }\textbf {\bibinfo
  {volume} {79}},\ \bibinfo {pages} {094504} (\bibinfo {year}
  {2016})}\BibitemShut {NoStop}%
\bibitem [{\citenamefont {Lin}\ \emph {et~al.}(2014)\citenamefont {Lin},
  \citenamefont {Xie}, \citenamefont {Osakada}, \citenamefont {Cui},\ and\
  \citenamefont {Cui}}]{lin14}%
  \BibitemOpen
  \bibfield  {author} {\bibinfo {author} {\bibfnamefont {Z.~C.}\ \bibnamefont
  {Lin}}, \bibinfo {author} {\bibfnamefont {C.}~\bibnamefont {Xie}}, \bibinfo
  {author} {\bibfnamefont {Y.}~\bibnamefont {Osakada}}, \bibinfo {author}
  {\bibfnamefont {Y.}~\bibnamefont {Cui}}, \ and\ \bibinfo {author}
  {\bibfnamefont {B.}~\bibnamefont {Cui}},\ }\bibfield  {title} {\enquote
  {\bibinfo {title} {Iridium oxide nanotube electrodes for sensitive and
  prolonged intracellular measurement of action potentials},}\ }\href {\doibase
  10.1038/ncomms4206} {\bibfield  {journal} {\bibinfo  {journal} {Nat.
  Commun.}\ }\textbf {\bibinfo {volume} {5}},\ \bibinfo {pages} {3206}
  (\bibinfo {year} {2014})}\BibitemShut {NoStop}%
\bibitem [{\citenamefont {Qiu}\ \emph {et~al.}(2015)\citenamefont {Qiu},
  \citenamefont {Hou}, \citenamefont {Kikkawa}, \citenamefont {Uchida},\ and\
  \citenamefont {Saitoh}}]{qiu15}%
  \BibitemOpen
  \bibfield  {author} {\bibinfo {author} {\bibfnamefont {Zh.}\ \bibnamefont
  {Qiu}}, \bibinfo {author} {\bibfnamefont {D.}~\bibnamefont {Hou}}, \bibinfo
  {author} {\bibfnamefont {T.}~\bibnamefont {Kikkawa}}, \bibinfo {author}
  {\bibfnamefont {K.}~\bibnamefont {Uchida}}, \ and\ \bibinfo {author}
  {\bibfnamefont {E.}~\bibnamefont {Saitoh}},\ }\bibfield  {title} {\enquote
  {\bibinfo {title} {All-oxide spin {Seebeck} effects},}\ }\href {\doibase
  10.7567/APEX.8.083001} {\bibfield  {journal} {\bibinfo  {journal} {Appl.
  Phys. Express}\ }\textbf {\bibinfo {volume} {8}},\ \bibinfo {pages} {083001}
  (\bibinfo {year} {2015})}\BibitemShut {NoStop}%
\bibitem [{\citenamefont {Hirsch}(1999)}]{hirsch99}%
  \BibitemOpen
  \bibfield  {author} {\bibinfo {author} {\bibfnamefont {J.~E.}\ \bibnamefont
  {Hirsch}},\ }\bibfield  {title} {\enquote {\bibinfo {title} {Spin hall
  effect},}\ }\href {\doibase 10.1103/PhysRevLett.83.1834} {\bibfield
  {journal} {\bibinfo  {journal} {Phys. Rev. Lett.}\ }\textbf {\bibinfo
  {volume} {83}},\ \bibinfo {pages} {1834--1837} (\bibinfo {year}
  {1999})}\BibitemShut {NoStop}%
\bibitem [{\citenamefont {Fujiwara}\ \emph {et~al.}(2013)\citenamefont
  {Fujiwara}, \citenamefont {Fukuma}, \citenamefont {Matsuno}, \citenamefont
  {Niimi}, \citenamefont {Otani},\ and\ \citenamefont {Takagi}}]{fuj13}%
  \BibitemOpen
  \bibfield  {author} {\bibinfo {author} {\bibfnamefont {K.}~\bibnamefont
  {Fujiwara}}, \bibinfo {author} {\bibfnamefont {Y.}~\bibnamefont {Fukuma}},
  \bibinfo {author} {\bibfnamefont {H.}~\bibnamefont {Matsuno}, \bibfnamefont
  {J.and~Idzuchi}}, \bibinfo {author} {\bibfnamefont {Y.}~\bibnamefont
  {Niimi}}, \bibinfo {author} {\bibfnamefont {Y.}~\bibnamefont {Otani}}, \ and\
  \bibinfo {author} {\bibfnamefont {H.}~\bibnamefont {Takagi}},\ }\bibfield
  {title} {\enquote {\bibinfo {title} {5d iridium oxide as a material for
  spin-current detection},}\ }\href {\doibase 10.1038/ncomms3893} {\bibfield
  {journal} {\bibinfo  {journal} {Nat. Comm.}\ }\textbf {\bibinfo {volume}
  {4}},\ \bibinfo {pages} {2893} (\bibinfo {year} {2013})}\BibitemShut
  {NoStop}%
\bibitem [{\citenamefont {Yan}\ \emph {et~al.}(2015)\citenamefont {Yan},
  \citenamefont {Ren}, \citenamefont {Xu}, \citenamefont {Xie}, \citenamefont
  {Tao}, \citenamefont {Choi}, \citenamefont {Lee}, \citenamefont {Choi},
  \citenamefont {Zhang},\ and\ \citenamefont {Feng}}]{yan15}%
  \BibitemOpen
  \bibfield  {author} {\bibinfo {author} {\bibfnamefont {Y.~J.}\ \bibnamefont
  {Yan}}, \bibinfo {author} {\bibfnamefont {M.~Q.}\ \bibnamefont {Ren}},
  \bibinfo {author} {\bibfnamefont {H.~C.}\ \bibnamefont {Xu}}, \bibinfo
  {author} {\bibfnamefont {B.~P.}\ \bibnamefont {Xie}}, \bibinfo {author}
  {\bibfnamefont {R.}~\bibnamefont {Tao}}, \bibinfo {author} {\bibfnamefont
  {H.~Y.}\ \bibnamefont {Choi}}, \bibinfo {author} {\bibfnamefont
  {N.}~\bibnamefont {Lee}}, \bibinfo {author} {\bibfnamefont {Y.~J.}\
  \bibnamefont {Choi}}, \bibinfo {author} {\bibfnamefont {T.}~\bibnamefont
  {Zhang}}, \ and\ \bibinfo {author} {\bibfnamefont {D.~L.}\ \bibnamefont
  {Feng}},\ }\bibfield  {title} {\enquote {\bibinfo {title} {{E}lectron-doped
  {Sr$_{2}$IrO$_{4}$}: An analogue of hole-doped cuprate superconductors
  demonstrated by scanning tunneling microscopy},}\ }\href {\doibase
  10.1103/PhysRevX.5.041018} {\bibfield  {journal} {\bibinfo  {journal} {Phys.
  Rev. X}\ }\textbf {\bibinfo {volume} {5}},\ \bibinfo {pages} {041018}
  (\bibinfo {year} {2015})}\BibitemShut {NoStop}%
\bibitem [{\citenamefont {Wang}\ and\ \citenamefont
  {Senthil}(2011)}]{wangsenth11}%
  \BibitemOpen
  \bibfield  {author} {\bibinfo {author} {\bibfnamefont {F.}~\bibnamefont
  {Wang}}\ and\ \bibinfo {author} {\bibfnamefont {T.}~\bibnamefont {Senthil}},\
  }\bibfield  {title} {\enquote {\bibinfo {title} {{Twisted Hubbard Model for
  {Sr$_{2}$IrO$_{4}$}: Magnetism and Possible High Temperature
  Superconductivity}},}\ }\href {\doibase 10.1103/PhysRevLett.106.136402}
  {\bibfield  {journal} {\bibinfo  {journal} {Phys. Rev. Lett.}\ }\textbf
  {\bibinfo {volume} {106}},\ \bibinfo {pages} {136402} (\bibinfo {year}
  {2011})}\BibitemShut {NoStop}%
\bibitem [{\citenamefont {R{\"{o}}hlsberger}(2004)}]{roe04}%
  \BibitemOpen
  \bibfield  {author} {\bibinfo {author} {\bibfnamefont {R.}~\bibnamefont
  {R{\"{o}}hlsberger}},\ }\href {\doibase 10.1007/b86125} {\emph {\bibinfo
  {title} {{Nuclear condensed matter physics with synchrotron radiation: basic
  principles, methodology and applications}}}},\ \bibinfo {series} {{Springer
  Tracts in Modern Physics}}, Vol.\ \bibinfo {volume} {208}\ (\bibinfo
  {publisher} {Springer},\ \bibinfo {address} {Heidelberg},\ \bibinfo {year}
  {2004})\BibitemShut {NoStop}%
\bibitem [{\citenamefont {Gerdau}\ and\ \citenamefont
  {DeWaard}(1999)}]{gerd99}%
  \BibitemOpen
  \bibinfo {editor} {\bibfnamefont {E.}~\bibnamefont {Gerdau}}\ and\ \bibinfo
  {editor} {\bibfnamefont {H.}~\bibnamefont {DeWaard}},\ eds.,\ \href {\doibase
  10.1023/A:1017073002352} {\emph {\bibinfo {title} {Nuclear resonant
  scattering of synchrotron radiation}}},\ Vol.\ \bibinfo {volume} {123-124}\
  (\bibinfo  {publisher} {Springer International Publishing},\ \bibinfo {year}
  {1999})\BibitemShut {NoStop}%
\bibitem [{\citenamefont {Wagner}\ and\ \citenamefont {Zahn}(1970)}]{wagn70}%
  \BibitemOpen
  \bibfield  {author} {\bibinfo {author} {\bibfnamefont {F.}~\bibnamefont
  {Wagner}}\ and\ \bibinfo {author} {\bibfnamefont {U.}~\bibnamefont {Zahn}},\
  }\bibfield  {title} {\enquote {\bibinfo {title} {M{\"o}ssbauer isomer shifts,
  hyperfine interactions, and magnetic hyperfine anomalies in compounds of
  iridium},}\ }\href {\doibase 10.1007/BF01396512} {\bibfield  {journal}
  {\bibinfo  {journal} {Z. Phys.}\ }\textbf {\bibinfo {volume} {233}},\
  \bibinfo {pages} {1--20} (\bibinfo {year} {1970})}\BibitemShut {NoStop}%
\bibitem [{\citenamefont {Ding}\ \emph {et~al.}(2016)\citenamefont {Ding},
  \citenamefont {Yang}, \citenamefont {Chen}, \citenamefont {Kim},
  \citenamefont {Han}, \citenamefont {Luo}, \citenamefont {Feng}, \citenamefont
  {Upton}, \citenamefont {Casa}, \citenamefont {Kim}, \citenamefont {Gog},
  \citenamefont {Zeng}, \citenamefont {Cao}, \citenamefont {Mao},\ and\
  \citenamefont {van Veenendaal}}]{ding16}%
  \BibitemOpen
  \bibfield  {author} {\bibinfo {author} {\bibfnamefont {Yang}\ \bibnamefont
  {Ding}}, \bibinfo {author} {\bibfnamefont {Liuxiang}\ \bibnamefont {Yang}},
  \bibinfo {author} {\bibfnamefont {Cheng-Chien}\ \bibnamefont {Chen}},
  \bibinfo {author} {\bibfnamefont {Heung-Sik}\ \bibnamefont {Kim}}, \bibinfo
  {author} {\bibfnamefont {Myung~Joon}\ \bibnamefont {Han}}, \bibinfo {author}
  {\bibfnamefont {Wei}\ \bibnamefont {Luo}}, \bibinfo {author} {\bibfnamefont
  {Zhenxing}\ \bibnamefont {Feng}}, \bibinfo {author} {\bibfnamefont {Mary}\
  \bibnamefont {Upton}}, \bibinfo {author} {\bibfnamefont {Diego}\ \bibnamefont
  {Casa}}, \bibinfo {author} {\bibfnamefont {Jungho}\ \bibnamefont {Kim}},
  \bibinfo {author} {\bibfnamefont {Thomas}\ \bibnamefont {Gog}}, \bibinfo
  {author} {\bibfnamefont {Zhidan}\ \bibnamefont {Zeng}}, \bibinfo {author}
  {\bibfnamefont {Gang}\ \bibnamefont {Cao}}, \bibinfo {author} {\bibfnamefont
  {Ho-kwang}\ \bibnamefont {Mao}}, \ and\ \bibinfo {author} {\bibfnamefont
  {Michel}\ \bibnamefont {van Veenendaal}},\ }\bibfield  {title} {\enquote
  {\bibinfo {title} {Pressure-induced confined metal from the mott insulator
  ${\mathrm{sr}}_{3}{\mathrm{ir}}_{2}{\mathrm{o}}_{7}$},}\ }\href {\doibase
  10.1103/PhysRevLett.116.216402} {\bibfield  {journal} {\bibinfo  {journal}
  {Phys. Rev. Lett.}\ }\textbf {\bibinfo {volume} {116}},\ \bibinfo {pages}
  {216402} (\bibinfo {year} {2016})}\BibitemShut {NoStop}%
\bibitem [{\citenamefont {Perlow}\ \emph {et~al.}(1969)\citenamefont {Perlow},
  \citenamefont {Henning}, \citenamefont {Olson},\ and\ \citenamefont
  {Goodman}}]{perl69}%
  \BibitemOpen
  \bibfield  {author} {\bibinfo {author} {\bibfnamefont {G.~J.}\ \bibnamefont
  {Perlow}}, \bibinfo {author} {\bibfnamefont {W.}~\bibnamefont {Henning}},
  \bibinfo {author} {\bibfnamefont {D.}~\bibnamefont {Olson}}, \ and\ \bibinfo
  {author} {\bibfnamefont {G.~L.}\ \bibnamefont {Goodman}},\ }\bibfield
  {title} {\enquote {\bibinfo {title} {Hyperfine anomaly in {$^{193}$Ir} by
  {M{\"o}ssbauer} effect, and its application to determination of the orbital
  part of hyperfine fields},}\ }\href {\doibase 10.1103/PhysRevLett.23.680}
  {\bibfield  {journal} {\bibinfo  {journal} {Phys. Rev. Lett.}\ }\textbf
  {\bibinfo {volume} {23}},\ \bibinfo {pages} {680--682} (\bibinfo {year}
  {1969})}\BibitemShut {NoStop}%
\bibitem [{\citenamefont {R{\"o}hlsberger}\ \emph {et~al.}(2001)\citenamefont
  {R{\"o}hlsberger}, \citenamefont {Bansmann}, \citenamefont {Senz},
  \citenamefont {Jonas}, \citenamefont {Bettac}, \citenamefont {Leupold},
  \citenamefont {R{\"u}ffer}, \citenamefont {Burkel},\ and\ \citenamefont
  {Meiwes-Broer}}]{roeh01}%
  \BibitemOpen
  \bibfield  {author} {\bibinfo {author} {\bibfnamefont {R.}~\bibnamefont
  {R{\"o}hlsberger}}, \bibinfo {author} {\bibfnamefont {J.}~\bibnamefont
  {Bansmann}}, \bibinfo {author} {\bibfnamefont {V.}~\bibnamefont {Senz}},
  \bibinfo {author} {\bibfnamefont {K.~L.}\ \bibnamefont {Jonas}}, \bibinfo
  {author} {\bibfnamefont {A.}~\bibnamefont {Bettac}}, \bibinfo {author}
  {\bibfnamefont {O.}~\bibnamefont {Leupold}}, \bibinfo {author} {\bibfnamefont
  {R.}~\bibnamefont {R{\"u}ffer}}, \bibinfo {author} {\bibfnamefont
  {E.}~\bibnamefont {Burkel}}, \ and\ \bibinfo {author} {\bibfnamefont {K.~H.}\
  \bibnamefont {Meiwes-Broer}},\ }\bibfield  {title} {\enquote {\bibinfo
  {title} {{Perpendicular Spin Orientation in Ultrasmall Fe Islands on
  W(110)}},}\ }\href {\doibase 10.1103/PhysRevLett.86.5597} {\bibfield
  {journal} {\bibinfo  {journal} {Phys. Rev. Lett.}\ }\textbf {\bibinfo
  {volume} {86}},\ \bibinfo {pages} {5597--5600} (\bibinfo {year}
  {2001})}\BibitemShut {NoStop}%
\bibitem [{\citenamefont {Stankov}\ \emph {et~al.}(2007)\citenamefont
  {Stankov}, \citenamefont {R{\"{o}}hlsberger}, \citenamefont {Slezak},
  \citenamefont {Sladecek}, \citenamefont {Sepiol}, \citenamefont {Vogl},
  \citenamefont {Chumakov}, \citenamefont {R{\"{u}}ffer}, \citenamefont
  {Spiridis}, \citenamefont {Lazewski}, \citenamefont {Parlinski},\ and\
  \citenamefont {Korecki}}]{stank07}%
  \BibitemOpen
  \bibfield  {author} {\bibinfo {author} {\bibfnamefont {S.}~\bibnamefont
  {Stankov}}, \bibinfo {author} {\bibfnamefont {R.}~\bibnamefont
  {R{\"{o}}hlsberger}}, \bibinfo {author} {\bibfnamefont {T.}~\bibnamefont
  {Slezak}}, \bibinfo {author} {\bibfnamefont {M.}~\bibnamefont {Sladecek}},
  \bibinfo {author} {\bibfnamefont {B.}~\bibnamefont {Sepiol}}, \bibinfo
  {author} {\bibfnamefont {G.}~\bibnamefont {Vogl}}, \bibinfo {author}
  {\bibfnamefont {A.~I.}\ \bibnamefont {Chumakov}}, \bibinfo {author}
  {\bibfnamefont {R.}~\bibnamefont {R{\"{u}}ffer}}, \bibinfo {author}
  {\bibfnamefont {N.}~\bibnamefont {Spiridis}}, \bibinfo {author}
  {\bibfnamefont {J.}~\bibnamefont {Lazewski}}, \bibinfo {author}
  {\bibfnamefont {K.}~\bibnamefont {Parlinski}}, \ and\ \bibinfo {author}
  {\bibfnamefont {J.}~\bibnamefont {Korecki}},\ }\bibfield  {title} {\enquote
  {\bibinfo {title} {{P}honons in {I}ron: {F}rom the {B}ulk to an {E}pitaxial
  {M}onolayer},}\ }\href {\doibase 10.1103/PhysRevLett.99.185501} {\bibfield
  {journal} {\bibinfo  {journal} {Phys. Rev. Lett.}\ }\textbf {\bibinfo
  {volume} {99}},\ \bibinfo {pages} {185501} (\bibinfo {year}
  {2007})}\BibitemShut {NoStop}%
\bibitem [{\citenamefont {Potapkin}\ \emph {et~al.}(2013)\citenamefont
  {Potapkin}, \citenamefont {McCammon}, \citenamefont {Glazyrin}, \citenamefont
  {Kantor}, \citenamefont {Kupenko}, \citenamefont {Prescher}, \citenamefont
  {Sinmyo}, \citenamefont {Smirnov}, \citenamefont {Chumakov},\ and\
  \citenamefont {R{\"u}ffer}}]{potapk13}%
  \BibitemOpen
  \bibfield  {author} {\bibinfo {author} {\bibfnamefont {V.}~\bibnamefont
  {Potapkin}}, \bibinfo {author} {\bibfnamefont {C.}~\bibnamefont {McCammon}},
  \bibinfo {author} {\bibfnamefont {K.}~\bibnamefont {Glazyrin}}, \bibinfo
  {author} {\bibfnamefont {A.}~\bibnamefont {Kantor}}, \bibinfo {author}
  {\bibfnamefont {I.}~\bibnamefont {Kupenko}}, \bibinfo {author} {\bibfnamefont
  {C.}~\bibnamefont {Prescher}}, \bibinfo {author} {\bibfnamefont
  {R.}~\bibnamefont {Sinmyo}}, \bibinfo {author} {\bibfnamefont {G.V.}\
  \bibnamefont {Smirnov}}, \bibinfo {author} {\bibfnamefont {A.I.}\
  \bibnamefont {Chumakov}}, \ and\ \bibinfo {author} {\bibfnamefont
  {R.}~\bibnamefont {R{\"u}ffer}},\ }\bibfield  {title} {\enquote {\bibinfo
  {title} {{Effect of iron oxidation state on the electrical conductivity of
  the Earth's lower mantle}},}\ }\href {\doibase 10.1038/ncomms2436} {\bibfield
   {journal} {\bibinfo  {journal} {Nat. Commun.}\ }\textbf {\bibinfo {volume}
  {4}},\ \bibinfo {pages} {1427} (\bibinfo {year} {2013})}\BibitemShut
  {NoStop}%
\bibitem [{\citenamefont {web site}()}]{p01c}%
  \BibitemOpen
  \bibfield  {author} {\bibinfo {author} {\bibfnamefont {Dynamics
  Beamline~P01}\ \bibnamefont {web site}},\ }\href
  {http://photon-science.desy.de/facilities} {\bibinfo  {journal}
  {http://photon-science.desy.de/facilities}\ }\BibitemShut {NoStop}%
\bibitem [{Sup()}]{SuppMat}%
  \BibitemOpen
\bibfield  {journal} {  }\href@noop {} {}\bibinfo {note} {See Supplemental
  Material at (URL will be inserted by publisher) for details on x-ray filter,
  APD detector array, beam cleaning in the PETRA ring, sensitivity of NRS to
  direction of hyperfine fields in iridates, and on alignment of the
  {Sr$_{2}$IrO$_{4}$} single crystals which includes Refs. \cite{serg07,
  gerd99, roe04, keil, klute, nawr, Sturh00, Sturh94, sung16,
  gret16}}\BibitemShut {NoStop}%
\bibitem [{\citenamefont {Baron}\ \emph {et~al.}(2006)\citenamefont {Baron},
  \citenamefont {Kishimoto}, \citenamefont {Morse},\ and\ \citenamefont
  {Rigal}}]{bar06}%
  \BibitemOpen
  \bibfield  {author} {\bibinfo {author} {\bibfnamefont {Alfred Q.~R.}\
  \bibnamefont {Baron}}, \bibinfo {author} {\bibfnamefont {Shunji}\
  \bibnamefont {Kishimoto}}, \bibinfo {author} {\bibfnamefont {John}\
  \bibnamefont {Morse}}, \ and\ \bibinfo {author} {\bibfnamefont {Jean-Marie}\
  \bibnamefont {Rigal}},\ }\bibfield  {title} {\enquote {\bibinfo {title}
  {{Silicon avalanche photodiodes for direct detection of X-rays}},}\ }\href
  {\doibase 10.1107/S090904950503431X} {\bibfield  {journal} {\bibinfo
  {journal} {Journal of Synchrotron Radiation}\ }\textbf {\bibinfo {volume}
  {13}},\ \bibinfo {pages} {131--142} (\bibinfo {year} {2006})}\BibitemShut
  {NoStop}%
\bibitem [{\citenamefont {Sergueev}\ \emph {et~al.}(2007)\citenamefont
  {Sergueev}, \citenamefont {Chumakov}, \citenamefont {Beaume-Dang},
  \citenamefont {R{\"u}ffer}, \citenamefont {Strohm},\ and\ \citenamefont {van
  B{\"u}rck}}]{serg07}%
  \BibitemOpen
  \bibfield  {author} {\bibinfo {author} {\bibfnamefont {I.}~\bibnamefont
  {Sergueev}}, \bibinfo {author} {\bibfnamefont {A.~I.}\ \bibnamefont
  {Chumakov}}, \bibinfo {author} {\bibfnamefont {T.~H.~Deschaux}\ \bibnamefont
  {Beaume-Dang}}, \bibinfo {author} {\bibfnamefont {R.}~\bibnamefont
  {R{\"u}ffer}}, \bibinfo {author} {\bibfnamefont {C.}~\bibnamefont {Strohm}},
  \ and\ \bibinfo {author} {\bibfnamefont {U.}~\bibnamefont {van B{\"u}rck}},\
  }\bibfield  {title} {\enquote {\bibinfo {title} {Nuclear forward scattering
  for high energy {M\"ossbauer} transitions},}\ }\href {\doibase
  10.1103/PhysRevLett.99.097601} {\bibfield  {journal} {\bibinfo  {journal}
  {Phys. Rev. Lett.}\ }\textbf {\bibinfo {volume} {99}},\ \bibinfo {pages}
  {097601} (\bibinfo {year} {2007})}\BibitemShut {NoStop}%
\bibitem [{\citenamefont {Achterberg}\ \emph {et~al.}(2006)\citenamefont
  {Achterberg}, \citenamefont {Capurro}, \citenamefont {Marti}, \citenamefont
  {Vanin},\ and\ \citenamefont {Castro}}]{nudat06}%
  \BibitemOpen
  \bibfield  {author} {\bibinfo {author} {\bibfnamefont {E.}~\bibnamefont
  {Achterberg}}, \bibinfo {author} {\bibfnamefont {O.A.}\ \bibnamefont
  {Capurro}}, \bibinfo {author} {\bibfnamefont {G.V.}\ \bibnamefont {Marti}},
  \bibinfo {author} {\bibfnamefont {V.R.}\ \bibnamefont {Vanin}}, \ and\
  \bibinfo {author} {\bibfnamefont {R.M.}\ \bibnamefont {Castro}},\ }\bibfield
  {title} {\enquote {\bibinfo {title} {{N}uclear {D}ata {S}heets for {A} =
  193},}\ }\href {\doibase http://dx.doi.org/10.1016/j.nds.2005.12.001}
  {\bibfield  {journal} {\bibinfo  {journal} {Nuclear Data Sheets}\ }\textbf
  {\bibinfo {volume} {107}},\ \bibinfo {pages} {1--224} (\bibinfo {year}
  {2006})}\BibitemShut {NoStop}%
\bibitem [{\citenamefont {Kishimoto}\ \emph {et~al.}(2005)\citenamefont
  {Kishimoto}, \citenamefont {Yoda}, \citenamefont {Kobayashi}, \citenamefont
  {Kitao}, \citenamefont {Haruki},\ and\ \citenamefont {Seto}}]{kishi05}%
  \BibitemOpen
  \bibfield  {author} {\bibinfo {author} {\bibfnamefont {S.}~\bibnamefont
  {Kishimoto}}, \bibinfo {author} {\bibfnamefont {Y.}~\bibnamefont {Yoda}},
  \bibinfo {author} {\bibfnamefont {Y.}~\bibnamefont {Kobayashi}}, \bibinfo
  {author} {\bibfnamefont {S.}~\bibnamefont {Kitao}}, \bibinfo {author}
  {\bibfnamefont {R.}~\bibnamefont {Haruki}}, \ and\ \bibinfo {author}
  {\bibfnamefont {M.}~\bibnamefont {Seto}},\ }\bibfield  {title} {\enquote
  {\bibinfo {title} {Evidence for nuclear excitation by electron transition on
  {$^{193}$Ir} and its probability},}\ }\href {\doibase
  http://dx.doi.org/10.1016/j.nuclphysa.2004.10.016} {\bibfield  {journal}
  {\bibinfo  {journal} {Nucl. Phys. A}\ }\textbf {\bibinfo {volume} {748}},\
  \bibinfo {pages} {3--11} (\bibinfo {year} {2005})}\BibitemShut {NoStop}%
\bibitem [{\citenamefont {Wagner}(1983)}]{wagn83}%
  \BibitemOpen
  \bibfield  {author} {\bibinfo {author} {\bibfnamefont {F.~E.}\ \bibnamefont
  {Wagner}},\ }\bibfield  {title} {\enquote {\bibinfo {title} {M{\"o}ssbauer
  spectroscopy with {$^{191,193}$Ir}},}\ }\href {\doibase 10.1007/BF01027249}
  {\bibfield  {journal} {\bibinfo  {journal} {Hyperfine Interact.}\ }\textbf
  {\bibinfo {volume} {13}},\ \bibinfo {pages} {149--173} (\bibinfo {year}
  {1983})}\BibitemShut {NoStop}%
\bibitem [{\citenamefont {Hannon}\ and\ \citenamefont
  {Trammell}(1999)}]{hann99}%
  \BibitemOpen
  \bibfield  {author} {\bibinfo {author} {\bibfnamefont {J.P.}\ \bibnamefont
  {Hannon}}\ and\ \bibinfo {author} {\bibfnamefont {G.T.}\ \bibnamefont
  {Trammell}},\ }\bibfield  {title} {\enquote {\bibinfo {title} {{Coherent
  $\gamma$-ray optics}},}\ }\href {\doibase 10.1023/A:1017011621007} {\bibfield
   {journal} {\bibinfo  {journal} {Hyperfine Interact.}\ }\textbf {\bibinfo
  {volume} {123}},\ \bibinfo {pages} {127--274} (\bibinfo {year}
  {1999})}\BibitemShut {NoStop}%
\bibitem [{\citenamefont {Kandiner}(2013)}]{kand13}%
  \BibitemOpen
  \bibfield  {author} {\bibinfo {author} {\bibfnamefont {H.J.}\ \bibnamefont
  {Kandiner}},\ }\href {https://books.google.de/books?id=mImABwAAQBAJ} {\emph
  {\bibinfo {title} {Iridium}}},\ {Gmelin Handbook of Inorganic and
  Organometallic Chemistry - 8th edition}\ (\bibinfo  {publisher} {Springer
  Berlin Heidelberg},\ \bibinfo {year} {2013})\BibitemShut {NoStop}%
\bibitem [{\citenamefont {G{\"{u}}tlich}\ \emph {et~al.}(2011)\citenamefont
  {G{\"{u}}tlich}, \citenamefont {Bill},\ and\ \citenamefont
  {Trautwein}}]{Guetl}%
  \BibitemOpen
  \bibfield  {author} {\bibinfo {author} {\bibfnamefont {Ph.}\ \bibnamefont
  {G{\"{u}}tlich}}, \bibinfo {author} {\bibfnamefont {E.}~\bibnamefont {Bill}},
  \ and\ \bibinfo {author} {\bibfnamefont {A.~X.}\ \bibnamefont {Trautwein}},\
  }\href {\doibase 10.1007/978-3-540-88428-6} {\emph {\bibinfo {title}
  {M{\"{o}}ssbauer Spectroscopy and Transition Metal Chemistry Fundamentals and
  Applications}}}\ (\bibinfo  {publisher} {Springer},\ \bibinfo {address}
  {Heidelberg},\ \bibinfo {year} {2011})\BibitemShut {NoStop}%
\bibitem [{\citenamefont {Steiner}\ \emph {et~al.}(1969)\citenamefont
  {Steiner}, \citenamefont {Gerdau}, \citenamefont {Hautsch},\ and\
  \citenamefont {Steenken}}]{stein69}%
  \BibitemOpen
  \bibfield  {author} {\bibinfo {author} {\bibfnamefont {P.}~\bibnamefont
  {Steiner}}, \bibinfo {author} {\bibfnamefont {E.}~\bibnamefont {Gerdau}},
  \bibinfo {author} {\bibfnamefont {W.}~\bibnamefont {Hautsch}}, \ and\
  \bibinfo {author} {\bibfnamefont {D.}~\bibnamefont {Steenken}},\ }\bibfield
  {title} {\enquote {\bibinfo {title} {Determination of the mean life of some
  excited nuclear states by {M{\"o}ssbauer} experiments},}\ }\href {\doibase
  10.1007/BF01392179} {\bibfield  {journal} {\bibinfo  {journal} {Z. Phys. A}\
  }\textbf {\bibinfo {volume} {221}},\ \bibinfo {pages} {281--290} (\bibinfo
  {year} {1969})}\BibitemShut {NoStop}%
\bibitem [{\citenamefont {Sturhahn}(2000)}]{Sturh00}%
  \BibitemOpen
  \bibfield  {author} {\bibinfo {author} {\bibfnamefont {W.}~\bibnamefont
  {Sturhahn}},\ }\bibfield  {title} {\enquote {\bibinfo {title} {{CONUSS} and
  {PHOENIX}: Evaluation of nuclear resonant scattering data},}\ }\href
  {\doibase 10.1023/A:1012681503686} {\bibfield  {journal} {\bibinfo  {journal}
  {Hyperfine Interact.}\ }\textbf {\bibinfo {volume} {125}},\ \bibinfo {pages}
  {149--172} (\bibinfo {year} {2000})}\BibitemShut {NoStop}%
\bibitem [{\citenamefont {Sturhahn}\ and\ \citenamefont
  {Gerdau}(1994)}]{Sturh94}%
  \BibitemOpen
  \bibfield  {author} {\bibinfo {author} {\bibfnamefont {W.}~\bibnamefont
  {Sturhahn}}\ and\ \bibinfo {author} {\bibfnamefont {E.}~\bibnamefont
  {Gerdau}},\ }\bibfield  {title} {\enquote {\bibinfo {title} {Evaluation of
  time$-$differential measurements of nuclear$-$resonance scattering of
  x$-$rays},}\ }\href {\doibase 10.1103/PhysRevB.49.9285} {\bibfield  {journal}
  {\bibinfo  {journal} {Phys. Rev. B}\ }\textbf {\bibinfo {volume} {49}},\
  \bibinfo {pages} {9285--9294} (\bibinfo {year} {1994})}\BibitemShut {NoStop}%
\bibitem [{\citenamefont {Atzmony}\ \emph {et~al.}(1967)\citenamefont
  {Atzmony}, \citenamefont {Bauminger}, \citenamefont {Lebenbaum},
  \citenamefont {Mustachi}, \citenamefont {Ofer},\ and\ \citenamefont
  {Wernick}}]{atzm67}%
  \BibitemOpen
  \bibfield  {author} {\bibinfo {author} {\bibfnamefont {U.}~\bibnamefont
  {Atzmony}}, \bibinfo {author} {\bibfnamefont {E.~R.}\ \bibnamefont
  {Bauminger}}, \bibinfo {author} {\bibfnamefont {D.}~\bibnamefont
  {Lebenbaum}}, \bibinfo {author} {\bibfnamefont {A.}~\bibnamefont {Mustachi}},
  \bibinfo {author} {\bibfnamefont {S.}~\bibnamefont {Ofer}}, \ and\ \bibinfo
  {author} {\bibfnamefont {J.~H.}\ \bibnamefont {Wernick}},\ }\bibfield
  {title} {\enquote {\bibinfo {title} {{M{\"{o}}ssbauer} effect in
  {$^{193}${Ir}} in intermetallic compounds and salts of iridium},}\ }\href
  {\doibase 10.1103/PhysRev.163.314} {\bibfield  {journal} {\bibinfo  {journal}
  {Phys. Rev.}\ }\textbf {\bibinfo {volume} {163}},\ \bibinfo {pages}
  {314--323} (\bibinfo {year} {1967})}\BibitemShut {NoStop}%
\bibitem [{\citenamefont {Bessas}\ \emph {et~al.}(2014)\citenamefont {Bessas},
  \citenamefont {Merkel}, \citenamefont {Chumakov}, \citenamefont
  {R{\"{u}}ffer}, \citenamefont {Hermann}, \citenamefont {Sergueev},
  \citenamefont {Mahmoud}, \citenamefont {Klobes}, \citenamefont {McGuire},
  \citenamefont {Sougrati},\ and\ \citenamefont {Stievano}}]{bess14}%
  \BibitemOpen
  \bibfield  {author} {\bibinfo {author} {\bibfnamefont {D.}~\bibnamefont
  {Bessas}}, \bibinfo {author} {\bibfnamefont {D.~G.}\ \bibnamefont {Merkel}},
  \bibinfo {author} {\bibfnamefont {A.~I.}\ \bibnamefont {Chumakov}}, \bibinfo
  {author} {\bibfnamefont {R.}~\bibnamefont {R{\"{u}}ffer}}, \bibinfo {author}
  {\bibfnamefont {R.~P.}\ \bibnamefont {Hermann}}, \bibinfo {author}
  {\bibfnamefont {I.}~\bibnamefont {Sergueev}}, \bibinfo {author}
  {\bibfnamefont {A.}~\bibnamefont {Mahmoud}}, \bibinfo {author} {\bibfnamefont
  {B.}~\bibnamefont {Klobes}}, \bibinfo {author} {\bibfnamefont {M.~A.}\
  \bibnamefont {McGuire}}, \bibinfo {author} {\bibfnamefont {M.~T.}\
  \bibnamefont {Sougrati}}, \ and\ \bibinfo {author} {\bibfnamefont
  {L.}~\bibnamefont {Stievano}},\ }\bibfield  {title} {\enquote {\bibinfo
  {title} {Nuclear forward scattering of synchrotron radiation by
  {$^{99}${Ru}}},}\ }\href {\doibase 10.1103/PhysRevLett.113.147601} {\bibfield
   {journal} {\bibinfo  {journal} {Phys. Rev. Lett.}\ }\textbf {\bibinfo
  {volume} {113}},\ \bibinfo {pages} {147601} (\bibinfo {year}
  {2014})}\BibitemShut {NoStop}%
\bibitem [{\citenamefont {Raghavan}\ \emph {et~al.}(1976)\citenamefont
  {Raghavan}, \citenamefont {Kaufmann}, \citenamefont {Raghavan}, \citenamefont
  {Ansaldo},\ and\ \citenamefont {Naumann}}]{ragh76}%
  \BibitemOpen
  \bibfield  {author} {\bibinfo {author} {\bibfnamefont {P.}~\bibnamefont
  {Raghavan}}, \bibinfo {author} {\bibfnamefont {E.~N.}\ \bibnamefont
  {Kaufmann}}, \bibinfo {author} {\bibfnamefont {R.~S.}\ \bibnamefont
  {Raghavan}}, \bibinfo {author} {\bibfnamefont {E.~J.}\ \bibnamefont
  {Ansaldo}}, \ and\ \bibinfo {author} {\bibfnamefont {R.~A.}\ \bibnamefont
  {Naumann}},\ }\bibfield  {title} {\enquote {\bibinfo {title} {Sign and
  magnitude of the quadrupole interaction of {$^{111}${Cd}} in noncubic metals:
  Universal correlation of ionic and electronic field gradients},}\ }\href
  {\doibase 10.1103/PhysRevB.13.2835} {\bibfield  {journal} {\bibinfo
  {journal} {Phys. Rev. B}\ }\textbf {\bibinfo {volume} {13}},\ \bibinfo
  {pages} {2835--2847} (\bibinfo {year} {1976})}\BibitemShut {NoStop}%
\bibitem [{\citenamefont {Shenoy}\ and\ \citenamefont {Wagner}(1984)}]{shen84}%
  \BibitemOpen
  \bibfield  {author} {\bibinfo {author} {\bibfnamefont {G.~K.}\ \bibnamefont
  {Shenoy}}\ and\ \bibinfo {author} {\bibfnamefont {F.E.}\ \bibnamefont
  {Wagner}},\ }\enquote {\bibinfo {title} {{M{\"o}ssbauer-Effect Isomer
  Shifts}},}\ in\ \href {\doibase 10.1007/978-1-4899-0462-1_5} {\emph {\bibinfo
  {booktitle} {M{\"o}ssbauer Spectroscopy Applied to Inorganic Chemistry}}},\
  \bibinfo {editor} {edited by\ \bibinfo {editor} {\bibfnamefont {G.~J.}\
  \bibnamefont {Long}}}\ (\bibinfo  {publisher} {Springer US},\ \bibinfo
  {address} {Boston, MA},\ \bibinfo {year} {1984})\ pp.\ \bibinfo {pages}
  {57--76}\BibitemShut {NoStop}%
\bibitem [{\citenamefont {M{\"o}ssbauer}\ \emph {et~al.}(1971)\citenamefont
  {M{\"o}ssbauer}, \citenamefont {Lengsfeld}, \citenamefont {von Lieres},
  \citenamefont {Potzel}, \citenamefont {Teschner}, \citenamefont {Wagner},\
  and\ \citenamefont {Kaindl}}]{moessb71}%
  \BibitemOpen
  \bibfield  {author} {\bibinfo {author} {\bibfnamefont {R.L.}\ \bibnamefont
  {M{\"o}ssbauer}}, \bibinfo {author} {\bibfnamefont {M.}~\bibnamefont
  {Lengsfeld}}, \bibinfo {author} {\bibfnamefont {W.}~\bibnamefont {von
  Lieres}}, \bibinfo {author} {\bibfnamefont {W.}~\bibnamefont {Potzel}},
  \bibinfo {author} {\bibfnamefont {P.}~\bibnamefont {Teschner}}, \bibinfo
  {author} {\bibfnamefont {F.E.}\ \bibnamefont {Wagner}}, \ and\ \bibinfo
  {author} {\bibfnamefont {G.}~\bibnamefont {Kaindl}},\ }\bibfield  {title}
  {\enquote {\bibinfo {title} {Nuclear gamma resonance study of the {Ir$-$Fe}
  and {Ir$-$Ni} alloy systems},}\ }\href {\doibase 10.1515/zna-1971-0303}
  {\bibfield  {journal} {\bibinfo  {journal} {Z. Naturforsch. A}\ }\textbf
  {\bibinfo {volume} {26}},\ \bibinfo {pages} {343} (\bibinfo {year}
  {1971})}\BibitemShut {NoStop}%
\bibitem [{\citenamefont {Salomon}\ and\ \citenamefont
  {Shirley}(1974)}]{salom74}%
  \BibitemOpen
  \bibfield  {author} {\bibinfo {author} {\bibfnamefont {D.}~\bibnamefont
  {Salomon}}\ and\ \bibinfo {author} {\bibfnamefont {D.~A.}\ \bibnamefont
  {Shirley}},\ }\bibfield  {title} {\enquote {\bibinfo {title} {Quadrupole
  coupling at {$^{193}$Ir} nuclei in iron},}\ }\href {\doibase
  10.1103/PhysRevB.9.29} {\bibfield  {journal} {\bibinfo  {journal} {Phys. Rev.
  B}\ }\textbf {\bibinfo {volume} {9}},\ \bibinfo {pages} {29--31} (\bibinfo
  {year} {1974})}\BibitemShut {NoStop}%
\bibitem [{\citenamefont {Rao}(1975)}]{rao75}%
  \BibitemOpen
  \bibfield  {author} {\bibinfo {author} {\bibfnamefont {G.N.}\ \bibnamefont
  {Rao}},\ }\bibfield  {title} {\enquote {\bibinfo {title} {{Dilute-impurity
  hyperfine fields in Fe, Co, Ni, and Gd}},}\ }\href {\doibase
  http://dx.doi.org/10.1016/0092-640X(75)90016-9} {\bibfield  {journal}
  {\bibinfo  {journal} {{At. Data Nucl. Data Tables}}\ }\textbf {\bibinfo
  {volume} {15}},\ \bibinfo {pages} {553 -- 576} (\bibinfo {year}
  {1975})}\BibitemShut {NoStop}%
\bibitem [{\citenamefont {Zhao}\ \emph {et~al.}(2008)\citenamefont {Zhao},
  \citenamefont {Yang}, \citenamefont {Yu}, \citenamefont {Li}, \citenamefont
  {Yu}, \citenamefont {Fang}, \citenamefont {Chen},\ and\ \citenamefont
  {Jin}}]{zhao08}%
  \BibitemOpen
  \bibfield  {author} {\bibinfo {author} {\bibfnamefont {J.G.}\ \bibnamefont
  {Zhao}}, \bibinfo {author} {\bibfnamefont {L.X.}\ \bibnamefont {Yang}},
  \bibinfo {author} {\bibfnamefont {Y.}~\bibnamefont {Yu}}, \bibinfo {author}
  {\bibfnamefont {F.Y.}\ \bibnamefont {Li}}, \bibinfo {author} {\bibfnamefont
  {R.C.}\ \bibnamefont {Yu}}, \bibinfo {author} {\bibfnamefont
  {Z.}~\bibnamefont {Fang}}, \bibinfo {author} {\bibfnamefont {L.C.}\
  \bibnamefont {Chen}}, \ and\ \bibinfo {author} {\bibfnamefont {C.Q.}\
  \bibnamefont {Jin}},\ }\bibfield  {title} {\enquote {\bibinfo {title}
  {High-pressure synthesis of orthorhombic sriro$_3$ perovskite and its
  positive magnetoresistance},}\ }\href {\doibase 10.1063/1.2908879} {\bibfield
   {journal} {\bibinfo  {journal} {J. of Appl. Phys.}\ }\textbf {\bibinfo
  {volume} {103}},\ \bibinfo {pages} {103706} (\bibinfo {year}
  {2008})}\BibitemShut {NoStop}%
\bibitem [{\citenamefont {Ingalls}(1964)}]{inga64}%
  \BibitemOpen
  \bibfield  {author} {\bibinfo {author} {\bibfnamefont {R.}~\bibnamefont
  {Ingalls}},\ }\bibfield  {title} {\enquote {\bibinfo {title}
  {{Electric-Field} {Gradient} {Tensor} in {Ferrous Compounds}},}\ }\href
  {\doibase 10.1103/PhysRev.133.A787} {\bibfield  {journal} {\bibinfo
  {journal} {Phys. Rev.}\ }\textbf {\bibinfo {volume} {133}},\ \bibinfo {pages}
  {A787--A795} (\bibinfo {year} {1964})}\BibitemShut {NoStop}%
\bibitem [{\citenamefont {Nie}\ \emph {et~al.}(2015)\citenamefont {Nie},
  \citenamefont {King}, \citenamefont {Kim}, \citenamefont {Uchida},
  \citenamefont {Wei}, \citenamefont {Faeth}, \citenamefont {Ruf},
  \citenamefont {Ruff}, \citenamefont {Xie}, \citenamefont {Pan}, \citenamefont
  {Fennie}, \citenamefont {Schlom},\ and\ \citenamefont {Shen}}]{nie15}%
  \BibitemOpen
  \bibfield  {author} {\bibinfo {author} {\bibfnamefont {Y.~F.}\ \bibnamefont
  {Nie}}, \bibinfo {author} {\bibfnamefont {P.~D.~C.}\ \bibnamefont {King}},
  \bibinfo {author} {\bibfnamefont {C.~H.}\ \bibnamefont {Kim}}, \bibinfo
  {author} {\bibfnamefont {M.}~\bibnamefont {Uchida}}, \bibinfo {author}
  {\bibfnamefont {H.~I.}\ \bibnamefont {Wei}}, \bibinfo {author} {\bibfnamefont
  {B.~D.}\ \bibnamefont {Faeth}}, \bibinfo {author} {\bibfnamefont {J.~P.}\
  \bibnamefont {Ruf}}, \bibinfo {author} {\bibfnamefont {J.~P.~C.}\
  \bibnamefont {Ruff}}, \bibinfo {author} {\bibfnamefont {L.}~\bibnamefont
  {Xie}}, \bibinfo {author} {\bibfnamefont {X.}~\bibnamefont {Pan}}, \bibinfo
  {author} {\bibfnamefont {C.~J.}\ \bibnamefont {Fennie}}, \bibinfo {author}
  {\bibfnamefont {D.~G.}\ \bibnamefont {Schlom}}, \ and\ \bibinfo {author}
  {\bibfnamefont {K.~M.}\ \bibnamefont {Shen}},\ }\bibfield  {title} {\enquote
  {\bibinfo {title} {{Interplay of Spin-Orbit Interactions, Dimensionality, and
  Octahedral Rotations in Semimetallic SrIrO$_3$}},}\ }\href {\doibase
  10.1103/PhysRevLett.114.016401} {\bibfield  {journal} {\bibinfo  {journal}
  {Phys. Rev. Lett.}\ }\textbf {\bibinfo {volume} {114}},\ \bibinfo {pages}
  {016401} (\bibinfo {year} {2015})}\BibitemShut {NoStop}%
\bibitem [{\citenamefont {Ye}\ \emph {et~al.}(2013)\citenamefont {Ye},
  \citenamefont {Chi}, \citenamefont {Chakoumakos}, \citenamefont
  {Fernandez-Baca}, \citenamefont {Qi},\ and\ \citenamefont {Cao}}]{yefe13}%
  \BibitemOpen
  \bibfield  {author} {\bibinfo {author} {\bibfnamefont {Feng}\ \bibnamefont
  {Ye}}, \bibinfo {author} {\bibfnamefont {S.}~\bibnamefont {Chi}}, \bibinfo
  {author} {\bibfnamefont {B.~C.}\ \bibnamefont {Chakoumakos}}, \bibinfo
  {author} {\bibfnamefont {J.~A.}\ \bibnamefont {Fernandez-Baca}}, \bibinfo
  {author} {\bibfnamefont {T.}~\bibnamefont {Qi}}, \ and\ \bibinfo {author}
  {\bibfnamefont {G.}~\bibnamefont {Cao}},\ }\bibfield  {title} {\enquote
  {\bibinfo {title} {Magnetic and crystal structures of {Sr$_{2}$IrO$_{4}$}: A
  neutron diffraction study},}\ }\href {\doibase 10.1103/PhysRevB.87.140406}
  {\bibfield  {journal} {\bibinfo  {journal} {Phys. Rev. B}\ }\textbf {\bibinfo
  {volume} {87}},\ \bibinfo {pages} {140406} (\bibinfo {year}
  {2013})}\BibitemShut {NoStop}%
\bibitem [{\citenamefont {Sch{\"u}nemann}\ and\ \citenamefont
  {Winkler}(2000)}]{schuenem00}%
  \BibitemOpen
  \bibfield  {author} {\bibinfo {author} {\bibfnamefont {V.}~\bibnamefont
  {Sch{\"u}nemann}}\ and\ \bibinfo {author} {\bibfnamefont {H.}~\bibnamefont
  {Winkler}},\ }\bibfield  {title} {\enquote {\bibinfo {title} {Structure and
  dynamics of biomolecules studied by {M{\"o}ssbauer} spectroscopy},}\ }\href
  {\doibase 10.1088/0034-4885/63/3/202} {\bibfield  {journal} {\bibinfo
  {journal} {Rep. Prog. Phys.}\ }\textbf {\bibinfo {volume} {63}},\ \bibinfo
  {pages} {263} (\bibinfo {year} {2000})}\BibitemShut {NoStop}%
\bibitem [{\citenamefont {Gretarsson}\ \emph {et~al.}(2016)\citenamefont
  {Gretarsson}, \citenamefont {Sung}, \citenamefont {{H{\"o}ppner}},
  \citenamefont {Kim}, \citenamefont {Keimer},\ and\ \citenamefont
  {Le~Tacon}}]{gret16}%
  \BibitemOpen
  \bibfield  {author} {\bibinfo {author} {\bibfnamefont {H.}~\bibnamefont
  {Gretarsson}}, \bibinfo {author} {\bibfnamefont {N.~H.}\ \bibnamefont
  {Sung}}, \bibinfo {author} {\bibfnamefont {M.}~\bibnamefont {{H{\"o}ppner}}},
  \bibinfo {author} {\bibfnamefont {B.~J.}\ \bibnamefont {Kim}}, \bibinfo
  {author} {\bibfnamefont {B.}~\bibnamefont {Keimer}}, \ and\ \bibinfo {author}
  {\bibfnamefont {M.}~\bibnamefont {Le~Tacon}},\ }\bibfield  {title} {\enquote
  {\bibinfo {title} {Two-magnon raman scattering and pseudospin-lattice
  interactions in {Sr$_{2}$IrO$_{4}$} and {Sr$_{3}$Ir$_{2}$O$_{7}$}},}\ }\href
  {\doibase 10.1103/PhysRevLett.116.136401} {\bibfield  {journal} {\bibinfo
  {journal} {Phys. Rev. Lett.}\ }\textbf {\bibinfo {volume} {116}},\ \bibinfo
  {pages} {136401} (\bibinfo {year} {2016})}\BibitemShut {NoStop}%
\bibitem [{\citenamefont {Oosterhuis}\ and\ \citenamefont
  {Lang}(1969)}]{oost69}%
  \BibitemOpen
  \bibfield  {author} {\bibinfo {author} {\bibfnamefont {W.~T.}\ \bibnamefont
  {Oosterhuis}}\ and\ \bibinfo {author} {\bibfnamefont {G.}~\bibnamefont
  {Lang}},\ }\bibfield  {title} {\enquote {\bibinfo {title} {{M{\"o}ssbauer}
  effect in {K$_{3}$FeCN$_{6}$}},}\ }\href {\doibase 10.1103/PhysRev.178.439}
  {\bibfield  {journal} {\bibinfo  {journal} {Phys. Rev.}\ }\textbf {\bibinfo
  {volume} {178}},\ \bibinfo {pages} {439--456} (\bibinfo {year}
  {1969})}\BibitemShut {NoStop}%
\bibitem [{\citenamefont {Henning}(1967)}]{henn67}%
  \BibitemOpen
  \bibfield  {author} {\bibinfo {author} {\bibfnamefont {J.C.M.}\ \bibnamefont
  {Henning}},\ }\bibfield  {title} {\enquote {\bibinfo {title} {Covalency and
  hyperfine structure of $(3d)^5$ ions in crystal fields},}\ }\href {\doibase
  http://dx.doi.org/10.1016/0375-9601(67)90184-3} {\bibfield  {journal}
  {\bibinfo  {journal} {Phys. Lett. A}\ }\textbf {\bibinfo {volume} {24}},\
  \bibinfo {pages} {40 -- 42} (\bibinfo {year} {1967})}\BibitemShut {NoStop}%
\bibitem [{\citenamefont {Herlitschke}\ \emph {et~al.}(2014)\citenamefont
  {Herlitschke}, \citenamefont {Tchougreeff}, \citenamefont {Soudackov},
  \citenamefont {Klobes}, \citenamefont {Stork}, \citenamefont {Dronskowski},\
  and\ \citenamefont {Hermann}}]{herlit14}%
  \BibitemOpen
  \bibfield  {author} {\bibinfo {author} {\bibfnamefont {M.}~\bibnamefont
  {Herlitschke}}, \bibinfo {author} {\bibfnamefont {A.~L.}\ \bibnamefont
  {Tchougreeff}}, \bibinfo {author} {\bibfnamefont {A.~V.}\ \bibnamefont
  {Soudackov}}, \bibinfo {author} {\bibfnamefont {B.}~\bibnamefont {Klobes}},
  \bibinfo {author} {\bibfnamefont {L.}~\bibnamefont {Stork}}, \bibinfo
  {author} {\bibfnamefont {R.}~\bibnamefont {Dronskowski}}, \ and\ \bibinfo
  {author} {\bibfnamefont {R.~P.}\ \bibnamefont {Hermann}},\ }\bibfield
  {title} {\enquote {\bibinfo {title} {Magnetism and lattice dynamics of
  {FeNCN} compared to {FeO}},}\ }\href {\doibase 10.1039/C4NJ00097H} {\bibfield
   {journal} {\bibinfo  {journal} {New J. Chem.}\ }\textbf {\bibinfo {volume}
  {38}},\ \bibinfo {pages} {4670--4677} (\bibinfo {year} {2014})}\BibitemShut
  {NoStop}%
\bibitem [{\citenamefont {Bogdanov}\ \emph {et~al.}(2015)\citenamefont
  {Bogdanov}, \citenamefont {Katukuri}, \citenamefont {Romh{\'a}nyi},
  \citenamefont {Yushankhai}, \citenamefont {Kataev}, \citenamefont
  {B{\"u}chner}, \citenamefont {van~den Brink},\ and\ \citenamefont
  {Hozoi}}]{bogd15}%
  \BibitemOpen
  \bibfield  {author} {\bibinfo {author} {\bibfnamefont {N.~A.}\ \bibnamefont
  {Bogdanov}}, \bibinfo {author} {\bibfnamefont {V.~M.}\ \bibnamefont
  {Katukuri}}, \bibinfo {author} {\bibfnamefont {J.}~\bibnamefont
  {Romh{\'a}nyi}}, \bibinfo {author} {\bibfnamefont {V.}~\bibnamefont
  {Yushankhai}}, \bibinfo {author} {\bibfnamefont {V.}~\bibnamefont {Kataev}},
  \bibinfo {author} {\bibfnamefont {B.}~\bibnamefont {B{\"u}chner}}, \bibinfo
  {author} {\bibfnamefont {J.}~\bibnamefont {van~den Brink}}, \ and\ \bibinfo
  {author} {\bibfnamefont {L.}~\bibnamefont {Hozoi}},\ }\bibfield  {title}
  {\enquote {\bibinfo {title} {Orbital reconstruction in nonpolar tetravalent
  transition-metal oxide layers},}\ }\href {\doibase 10.1038/ncomms8306}
  {\bibfield  {journal} {\bibinfo  {journal} {Nat. Comm.}\ }\textbf {\bibinfo
  {volume} {6}},\ \bibinfo {pages} {73061--73069} (\bibinfo {year}
  {2015})}\BibitemShut {NoStop}%
\bibitem [{\citenamefont {Ishida}\ \emph {et~al.}(1997)\citenamefont {Ishida},
  \citenamefont {Kitaoka}, \citenamefont {Asayama}, \citenamefont {Ikeda},
  \citenamefont {Nishizaki}, \citenamefont {Maeno}, \citenamefont {Yoshida},\
  and\ \citenamefont {Fujita}}]{ishid97}%
  \BibitemOpen
  \bibfield  {author} {\bibinfo {author} {\bibfnamefont {K.}~\bibnamefont
  {Ishida}}, \bibinfo {author} {\bibfnamefont {Y.}~\bibnamefont {Kitaoka}},
  \bibinfo {author} {\bibfnamefont {K.}~\bibnamefont {Asayama}}, \bibinfo
  {author} {\bibfnamefont {S.}~\bibnamefont {Ikeda}}, \bibinfo {author}
  {\bibfnamefont {S.}~\bibnamefont {Nishizaki}}, \bibinfo {author}
  {\bibfnamefont {Y.}~\bibnamefont {Maeno}}, \bibinfo {author} {\bibfnamefont
  {K.}~\bibnamefont {Yoshida}}, \ and\ \bibinfo {author} {\bibfnamefont
  {T.}~\bibnamefont {Fujita}},\ }\bibfield  {title} {\enquote {\bibinfo {title}
  {Anisotropic pairing in superconducting {Sr$_{2}$RuO$_{4}$}: {Ru} {NMR} and
  {NQR} studies},}\ }\href {\doibase 10.1103/PhysRevB.56.R505} {\bibfield
  {journal} {\bibinfo  {journal} {Phys. Rev. B}\ }\textbf {\bibinfo {volume}
  {56}},\ \bibinfo {pages} {R505--R508} (\bibinfo {year} {1997})}\BibitemShut
  {NoStop}%
\bibitem [{\citenamefont {Nichols}\ \emph {et~al.}(2016)\citenamefont
  {Nichols}, \citenamefont {Gao}, \citenamefont {Lee}, \citenamefont {Meyer},
  \citenamefont {Freeland}, \citenamefont {Lauter}, \citenamefont {Yi},
  \citenamefont {Liu}, \citenamefont {Haskel},\ and\ \citenamefont
  {Petrie}}]{nich16}%
  \BibitemOpen
  \bibfield  {author} {\bibinfo {author} {\bibfnamefont {J.}~\bibnamefont
  {Nichols}}, \bibinfo {author} {\bibfnamefont {X.}~\bibnamefont {Gao}},
  \bibinfo {author} {\bibfnamefont {S.}~\bibnamefont {Lee}}, \bibinfo {author}
  {\bibfnamefont {T.~L.}\ \bibnamefont {Meyer}}, \bibinfo {author}
  {\bibfnamefont {J.~W.}\ \bibnamefont {Freeland}}, \bibinfo {author}
  {\bibfnamefont {V.}~\bibnamefont {Lauter}}, \bibinfo {author} {\bibfnamefont
  {D.}~\bibnamefont {Yi}}, \bibinfo {author} {\bibfnamefont {J.}~\bibnamefont
  {Liu}}, \bibinfo {author} {\bibfnamefont {D.}~\bibnamefont {Haskel}}, \ and\
  \bibinfo {author} {\bibfnamefont {J.~R.}\ \bibnamefont {Petrie}},\ }\bibfield
   {title} {\enquote {\bibinfo {title} {Emerging magnetism and anomalous {Hall}
  effect in iridate{-}manganite heterostructures},}\ }\href {\doibase
  10.1038/ncomms12721} {\bibfield  {journal} {\bibinfo  {journal} {Nat. Comm.}\
  }\textbf {\bibinfo {volume} {7}},\ \bibinfo {pages} {12721} (\bibinfo {year}
  {2016})}\BibitemShut {NoStop}%
\bibitem [{\citenamefont {Keil}\ and\ \citenamefont
  {Ehrlichmann}(2016)}]{keil}%
  \BibitemOpen
  \bibfield  {author} {\bibinfo {author} {\bibfnamefont {J.}~\bibnamefont
  {Keil}}\ and\ \bibinfo {author} {\bibfnamefont {H.}~\bibnamefont
  {Ehrlichmann}},\ }\bibfield  {title} {\enquote {\bibinfo {title} {{B}unch
  {P}urity {M}easurements at {PETRA} {III}},}\ \ }(\bibinfo {organization} {7th
  International Particle Accelerator Conference, Busan (Korea), 8 May$-$13
  May},\ \bibinfo {year} {2016})\ pp.\ \bibinfo {pages}
  {3434--3436}\BibitemShut {NoStop}%
\bibitem [{\citenamefont {Klute}\ \emph {et~al.}(2011)\citenamefont {Klute},
  \citenamefont {Balewski}, \citenamefont {Delfs}, \citenamefont {Duhme},
  \citenamefont {Ebert}, \citenamefont {Neumann},\ and\ \citenamefont
  {Obier}}]{klute}%
  \BibitemOpen
  \bibfield  {author} {\bibinfo {author} {\bibfnamefont {J.}~\bibnamefont
  {Klute}}, \bibinfo {author} {\bibfnamefont {K.}~\bibnamefont {Balewski}},
  \bibinfo {author} {\bibfnamefont {A.}~\bibnamefont {Delfs}}, \bibinfo
  {author} {\bibfnamefont {H.~T.}\ \bibnamefont {Duhme}}, \bibinfo {author}
  {\bibfnamefont {M.}~\bibnamefont {Ebert}}, \bibinfo {author} {\bibfnamefont
  {Ru.}\ \bibnamefont {Neumann}}, \ and\ \bibinfo {author} {\bibfnamefont
  {F.}~\bibnamefont {Obier}},\ }\bibfield  {title} {\enquote {\bibinfo {title}
  {{T}he {PETRA} {III} {M}ultibunch {F}eedback {S}ystem},}\ \ }(\bibinfo
  {organization} {10$^{th}$ European Workshop on Beam Diagnostics and
  Instrumentation for Particle Accelerators, Hamburg (Germany), 16 May$-$18
  May},\ \bibinfo {year} {2011})\ pp.\ \bibinfo {pages} {494--496}\BibitemShut
  {NoStop}%
\bibitem [{\citenamefont {Nawrocky}\ \emph {et~al.}(1993)\citenamefont
  {Nawrocky}, \citenamefont {Bergmann},\ and\ \citenamefont {Siddons}}]{nawr}%
  \BibitemOpen
  \bibfield  {author} {\bibinfo {author} {\bibfnamefont {R.J.}\ \bibnamefont
  {Nawrocky}}, \bibinfo {author} {\bibfnamefont {U.}~\bibnamefont {Bergmann}},
  \ and\ \bibinfo {author} {\bibfnamefont {D.P.}\ \bibnamefont {Siddons}},\
  }\bibfield  {title} {\enquote {\bibinfo {title} {A bunch killer for the nsls
  x-ray electron storage ring},}\ }in\ \href {\doibase 10.1109/PAC.1993.309249}
  {\emph {\bibinfo {booktitle} {Proceedings of International Conference on
  Particle Accelerators}}},\ Vol.~\bibinfo {volume} {3}\ (\bibinfo {year}
  {1993})\ pp.\ \bibinfo {pages} {2145--2147}\BibitemShut {NoStop}%
\bibitem [{\citenamefont {Sung}\ \emph {et~al.}(2016)\citenamefont {Sung},
  \citenamefont {Gretarsson}, \citenamefont {Proepper}, \citenamefont {Porras},
  \citenamefont {Le~Tacon}, \citenamefont {Boris}, \citenamefont {Keimer},\
  and\ \citenamefont {Kim}}]{sung16}%
  \BibitemOpen
  \bibfield  {author} {\bibinfo {author} {\bibfnamefont {N.H.}\ \bibnamefont
  {Sung}}, \bibinfo {author} {\bibfnamefont {H.}~\bibnamefont {Gretarsson}},
  \bibinfo {author} {\bibfnamefont {D.}~\bibnamefont {Proepper}}, \bibinfo
  {author} {\bibfnamefont {J.}~\bibnamefont {Porras}}, \bibinfo {author}
  {\bibfnamefont {M.}~\bibnamefont {Le~Tacon}}, \bibinfo {author}
  {\bibfnamefont {A.~V.}\ \bibnamefont {Boris}}, \bibinfo {author}
  {\bibfnamefont {B.}~\bibnamefont {Keimer}}, \ and\ \bibinfo {author}
  {\bibfnamefont {B.~J.}\ \bibnamefont {Kim}},\ }\bibfield  {title} {\enquote
  {\bibinfo {title} {Crystal growth and intrinsic magnetic behaviour of
  {Sr$_{2}$IrO$_{4}$}},}\ }\href {\doibase 10.1080/14786435.2015.1134835}
  {\bibfield  {journal} {\bibinfo  {journal} {Philos. Mag.}\ }\textbf {\bibinfo
  {volume} {96}},\ \bibinfo {pages} {413--426} (\bibinfo {year}
  {2016})}\BibitemShut {NoStop}%
\end{thebibliography}
%
\end{document}